\documentclass[useAMS,usenatbib,usegraphicx]{mn2e}
\usepackage[xcolor]{xcolor}

\def\kmsec{\mbox{km~s$^{\rm -1}$}}
\def\logg{\mbox{log~{\it g}}}

\def\teff{\mbox{$T_{\rm eff}$}}
\def\vt{\mbox{$v_{\rm t}$}}

\def\rpro{\mbox{$r$-process}}
\def\spro{\mbox{$s$-process}}
\def\ncap{\mbox{$n$-capture}}

\def\sga{\mbox{CS~22886--012}}
\def\sgb{\mbox{CS~22943--132}}
\def\sgc{\mbox{CS~22945--017}}
\def\sgd{\mbox{CS~22945--058}}
\def\sge{\mbox{CS~22958--052}}
\def\sgf{\mbox{CS~29529--054}}
\def\rga{\mbox{CS~22183--031}}
\def\rgb{\mbox{CS~22892--052}}
\def\rgc{\mbox{CS~22953--003}}
\def\rgd{\mbox{CS~31082--001}}
\def\hba{\mbox{CS~22875--029}}
\def\hbb{\mbox{CS~22882--001}}
\def\hbc{\mbox{CS~22888--047}}
\def\bd{\mbox{BD$+$17~3248}}

\def\hdone{\mbox{HD~115444}}
\def\apj{ApJ}
\def\aap{A\&A}
\def\aj{AJ}
\def\mnras{MNRAS}
\def\araa{ARA\&A}
\def\apjs{ApJS}
\def\apjl{ApJL}
\def\procspie{Proc.\ SPIE}

\def\pasp{PASP}
\def\pasj{PASJ}

\def\ssr{Space Sci.\ Rev.}

\title[Highly-$r$-process-enhanced stars]
{Nine New Metal-Poor Stars 
on the Subgiant and Red Horizontal Branches
with High Levels of $r$-process Enhancement\thanks{
This paper includes data gathered with the 6.5 meter 
Magellan Telescopes located at Las Campanas Observatory, Chile.
} }

\author[Roederer et al.]{%
Ian U.\ Roederer,$^{1}$\thanks{E-mail: iur@umich.edu}
John J.\ Cowan,$^{2}$
George W.\ Preston,$^{3}$ \newauthor
Stephen A. Shectman,$^{3}$
Christopher Sneden,$^{4}$
Ian B.\ Thompson$^{3}$\\
$^{1}$Department of Astronomy, University of Michigan,
1085 S.\ University Ave., Ann Arbor, MI 48109, USA\\
$^{2}$Homer L.\ Dodge Department of Physics and Astronomy,
University of Oklahoma, 440 W.\ Brooks St., Norman, OK 73019, USA\\
$^{3}$Carnegie Observatories, 
813 Santa Barbara St., Pasadena, CA 91101, USA\\
$^{4}$Department of Astronomy, University of Texas at Austin,
1 University Station C1400, Austin, TX 78712, USA
}

\begin{document}

\pagerange{\pageref{firstpage}--\pageref{lastpage}} 
\pubyear{2014}
\maketitle
\label{firstpage}


\begin{abstract}

We report the discovery of nine metal-poor stars
with high levels of $r$-process enhancement
($+$0.81~$\leq$~[Eu/Fe]~$\leq +$1.13),
including six subgiants and three stars 
on the
red horizontal branch.
We also analyze four previously-known 
$r$-process-enhanced
metal-poor red giants.
From this sample of 13~stars, we draw the following conclusions.
(1) High levels of $r$-process enhancement are found 
in a broad range of stellar evolutionary states,
reaffirming that this phenomenon is not
associated with a chemical peculiarity 
of red giant atmospheres.
(2) Only 1 of 10~stars observed at multiple epochs
shows radial velocity variations, 
reaffirming that
stars with high levels of $r$-process enhancement
are not preferentially found 
among binaries.
(3) Only 2 of the 13~stars are highly-enhanced in 
C and N, indicating that
there is no connection
between high levels of $r$-process enhancement and
high levels of C and N.
(4) The dispersions in [Sr/Ba] and [Sr/Eu] 
are larger than the dispersions in [Ba/Eu] and [Yb/Eu],
suggesting
that the 
elements below the second $r$-process peak
do not always scale with those in the rare earth domain,
even within the class of highly-$r$-process-enhanced stars.
(5) The light-element 
(12~$\leq Z \leq$~30)
abundances of
highly-$r$-process-enhanced stars are
indistinguishable from those with normal levels of $r$-process material
at the limit of our data, 3.5~per cent (0.015~dex) on average.
The nucleosynthetic sites responsible 
for the large $r$-process enhancements
did not produce any detectable light-element abundance signatures
distinct from normal core-collapse supernovae.

\end{abstract}


\section{Introduction}
\label{introduction}

Elements heavier than the iron group comprise a
miniscule fraction of all atoms by number or mass,
but they comprise approximately 70~per cent
of the stable or long-lived isotopes found naturally on earth.
The rapid neutron-capture process, also known as the \rpro, 
is one of the two general nucleosynthesis mechanisms
responsible for production of most isotopes
heavier than the iron group
\citep{burbidge57}.
Two decades ago, observations of the metal-poor star \rgb\
([Fe/H]~$= -$3.1)
revealed a surprising consistency between
its heavy-element abundance pattern and the
predicted \rpro\ component of the solar system distribution
\citep{sneden94,cowan95}.
Over the next decade, three other stars were found with
similar abundance patterns
(\hdone, \citealt{westin00};
\rgd, \citealt{hill02}; and  
\bd, \citealt{cowan02}), and since then
tens more have been identified
(e.g., \citealt{barklem05}).
The ratios of heavy elements to Fe (e.g., [Eu/Fe]) 
in many of these stars are an order of magnitude larger
than in the solar system,
and this is one of two defining characteristics of the
``$r$-II'' class of stars \citep{beers05}.
(The other defining characteristic is that
the star has [Ba/Eu]~$<$~0.)
The existence of this \rpro\ pattern 
in metal-poor stars nearly as old as the Universe itself
was not predicted by theory, and the stellar
sites responsible for \rpro\ nucleosynthesis
are still debated today.

Our use of the phrase \textit{\rpro\ nucleosynthesis}
refers to the generic process described by \citet{burbidge57}:\
``a large flux of neutrons becomes available
in a short time interval for addition to elements
of the iron group'' (p.\ 587).
That which produces
the predicted \rpro\ abundance pattern in solar system material
is the so-called \textit{main component} of the \rpro.
This is what \citeauthor{burbidge57}\ were trying to
reproduce with their analytic description,
even though it would be another 16~years before 
others would attempt to separate the \rpro\ contribution
explicitly from the slow neutron-capture (\spro) contribution
in the solar system total isotopic distribution
\citep{cameron73}.

The \citeauthor{burbidge57}\ definition is, however,
sufficiently flexible to account for
the ``incomplete \rpro\ synthesis'' 
(\citealt{truran02}, p.\ 1305)
abundance pattern, or so-called \textit{weak component},
found in many other metal-poor stars and exemplified by 
\mbox{HD~122563}
(e.g., \citealt{wallerstein63,sneden83,honda06}).
The patterns typical of the main and weak components
may represent the extremes of
a range of \rpro\ nucleosynthesis outcomes
that depend on the physical conditions at 
the time of nucleosynthesis.
They may also
represent two distinct processes
\citep{montes07}.

Abundance patterns intermediate between these
two extremes are found, and the full range
spans at least
$-$0.5~$\leq$~[Eu/Fe]~$\leq +$1.9
(e.g., Sneden, Cowan, \& Gallino \citeyear{sneden08}).
This can be illustrated, for example, with
the [Sr/Ba] ratio in $\approx$~100~stars as shown in 
in Figure~17 of \citet{cohen13}.
Alternatively, the same effect is shown
for many elements in 16~stars 
in Figure~11 of \citet{roederer10}.
\citet{francois07} and
\citet{siqueiramello14} also
find these intermediate values 
in their samples of metal-poor stars
with various levels of \rpro\ enhancement
(see their Figures~15 and 28, respectively).
Phenomenological chemical evolution 
models (e.g., \citealt{qian08,aoki13})
and physically-motivated \rpro\ calculations
(e.g., \citealt{kratz07,boyd12})
can reproduce the observed distributions.
These explanations, however, are not uniquely
capable of predicting the range of [Sr/Ba] ratios
in low metallicity stars
(e.g., \citealt{cescutti13}).

In the literature, references to \textit{the \rpro} 
frequently refer to the event producing
the main component pattern.
Observations
indicating that the \rpro\ is a rare event and those 
indicating the nearly-ubiquitous presence of \rpro\ material in stars
are not mutually exclusive.
Abundance patterns like that found in the star \rgb\ are
rare, and so presumably the nucleosynthesis events that
produce them are also rare.
The greater concentration of stars with
[Ba/Fe] and [Ba/Sr]~$<$~0 and low [C/Fe] ratios 
in Figure~7 of \citet{sneden08}
indicates that stars with these characteristics, 
like \mbox{HD~122563}, are not rare
(cf.\ \citealt{roederer13}).
Thus the \rpro\ events that give rise to these patterns
are probably not rare, either.

In this paper we present nine new members
of the class of stars highly-enhanced in \rpro\ material.
Until recent years the members of this class included
only evolved red giants,
as summarized by \citet{cowan11}.
This understandably led to concern that
perhaps the consistent \rpro\ pattern observed in some
metal-poor stars was an artifact of the
analysis or related to physics of the stellar photospheres
rather than nucleosynthesis.
The one exception is a faint main sequence dwarf identified
by \citet{aoki10}.
The nine stars 
presented here are subgiants
(including stars at the main-sequence turn-off, MSTO)
and the
field-star equivalents of red horizontal branch (RHB) stars.
These stars have been identified among the
metal-poor candidates presented by Beers, Preston, \& Shectman
(\citeyear{beers92})
and analyzed in detail by \citet{roederer14}.

The details of sample selection, observations, and a summary
of the analysis techniques
are presented in Sections~\ref{sample}, \ref{r14summary}, and
\ref{compare}.
We examine these stars for evidence of radial velocity variations
in Section~\ref{rv}.
Section~\ref{patterns} demonstrates the close matches
between the abundance patterns in these stars
and four previously-known highly-\rpro-enhanced giants,
including \rgb.
Sections~\ref{hb}, \ref{carbon}, and \ref{light} discuss
our results, and Section~\ref{summary} summarizes our findings.

Throughout this work we
adopt the standard definitions of elemental abundances and ratios.
For element X, the logarithmic abundance is defined
as the number of atoms of X per 10$^{12}$ hydrogen atoms,
$\log\epsilon$(X)~$\equiv \log_{10}(N_{\rm X}/N_{\rm H}) +$~12.0.
For elements X and Y, the logarithmic abundance ratio relative to the
solar ratio, denoted
[X/Y], is defined as $\log_{10} (N_{\rm X}/N_{\rm Y}) -
\log_{10} (N_{\rm X}/N_{\rm Y})_{\odot}$.
We adopt the solar abundances given by
\citet{asplund09} and listed in Table~13 of 
\citet{roederer14}.
Abundances or ratios denoted with the ionization state
indicate the total elemental abundance as derived from transitions of
that particular state 
after ionization corrections have been applied.
Abundance ratios for elements X and Y
compare the total abundances of X 
and Y derived from like ionization states;
i.e., neutrals with neutrals and ions with ions.

\section{Sample Selection}
\label{sample}

We draw our sample from the catalog of 313 metal-poor stars
observed and analyzed by \citet{roederer14}.
We adopt a traditional indicator of \rpro\ enrichment
to select stars highly-enhanced in \rpro\ material,
the [Eu/Fe] ratio.
The \citeauthor{roederer14}\ study included four red giants
previously identified as being highly-enhanced in \rpro\ material,
\rga, \rgb, \rgc, and \rgd.
Prior studies have generally confirmed that [Eu/Fe] $> +$1.0
in each of these stars.
As discussed in Section~\ref{compare}, 
the \citeauthor{roederer14}\ [Eu/Fe] ratios
are lower than those found by previous studies
of these four stars by an average of 0.22~dex ($\sigma =$~0.10).
Therefore, to remain consistent with the
$r$-II classification scheme, we consider stars with
[Eu/Fe]~$\ga +$0.8 as candidates for inclusion in this class.

\begin{figure}
\begin{center}
\includegraphics[angle=90,width=3.35in]{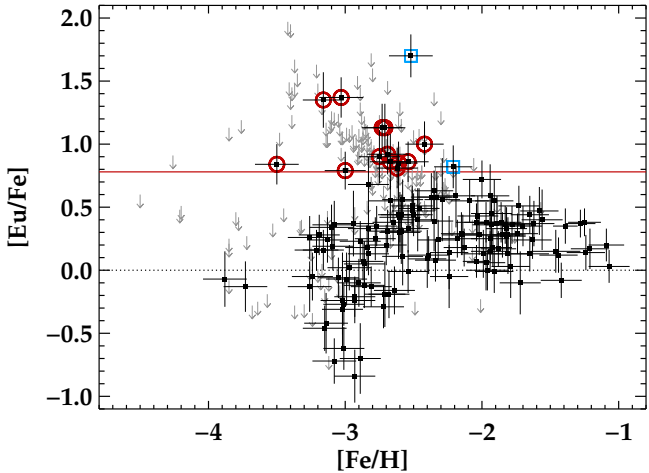}
\end{center}
\caption{
\label{euplot}
[Eu/Fe] ratios for all stars in the \citet{roederer14} sample.
The red line marks [Eu/Fe]~$= +$0.78, the 
lower limit for classifying a star as 
highly-enhanced in \rpro\ material.
Red circles mark stars examined in our study, and
the blue squares mark stars with high levels of [Eu/Fe]
resulting from \spro\ enrichment.
The dotted line marks the solar ratio.
}
\end{figure}

Figure~\ref{euplot} illustrates the [Eu/Fe] ratios for all stars
in the \citet{roederer14} sample.
The red line marks [Eu/Fe]~$= +$0.78.
Eu~\textsc{ii} lines are detected in 15~stars whose [Eu/Fe] ratios
exceed this limit.
Four of these stars are the previously-known giants 
mentioned above.
As we discuss in Section~\ref{patterns}, 
nine of the remaining Eu-rich 
stars have abundance patterns consistent 
with that found in previously-identified highly-\rpro-enhanced stars and the
calculated solar \rpro\ abundance pattern.
These stars are marked with large red circles in Figure~\ref{euplot}.
These nine stars include
six subgiants
(\sga, \sgb, \sgc, \sgd, \sge, \sgf)
and three field stars that occupy the
same region of the temperature-gravity diagram
as the RHB stars in globular clusters
(\hba, \hbb, \hbc).
\citet{preston06} presented abundances for a limited selection
of heavy elements in these three RHB stars.
Their results are similar to those presented here
(Section~\ref{compare}), 
but that study did not discuss the
high levels of \rpro\ enhancement found in these stars.

\begin{figure*}
\begin{center}
\includegraphics[angle=0,width=5.0in]{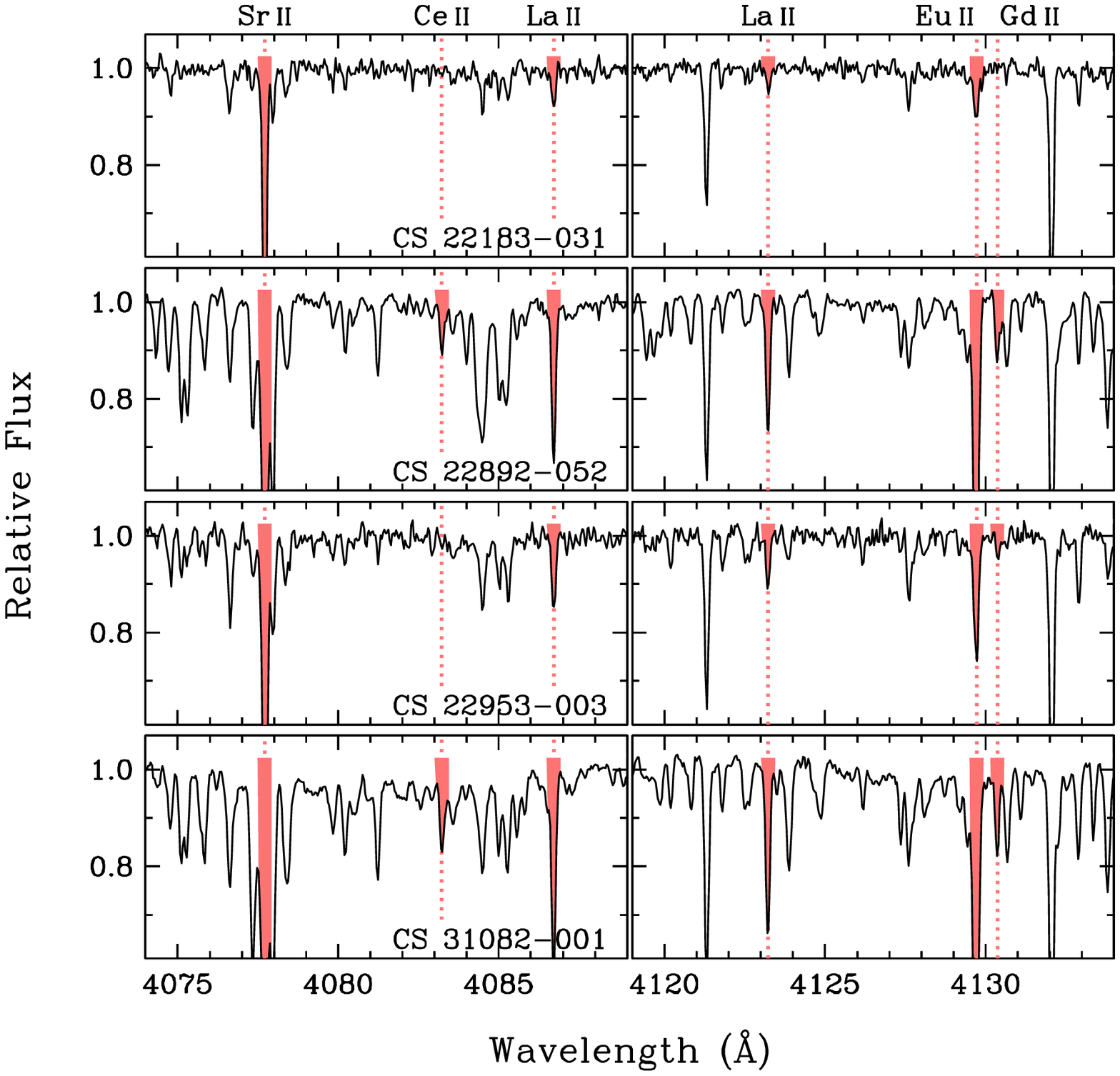}
\end{center}
\caption{
\label{rgbspecplot}
Spectra of the four red giant stars.
Shaded regions indicate lines used in the abundance analysis
of \ncap\ elements.
Species are indicated at the top.
}
\end{figure*}

\begin{figure*}
\begin{center}
\includegraphics[angle=0,width=5.0in]{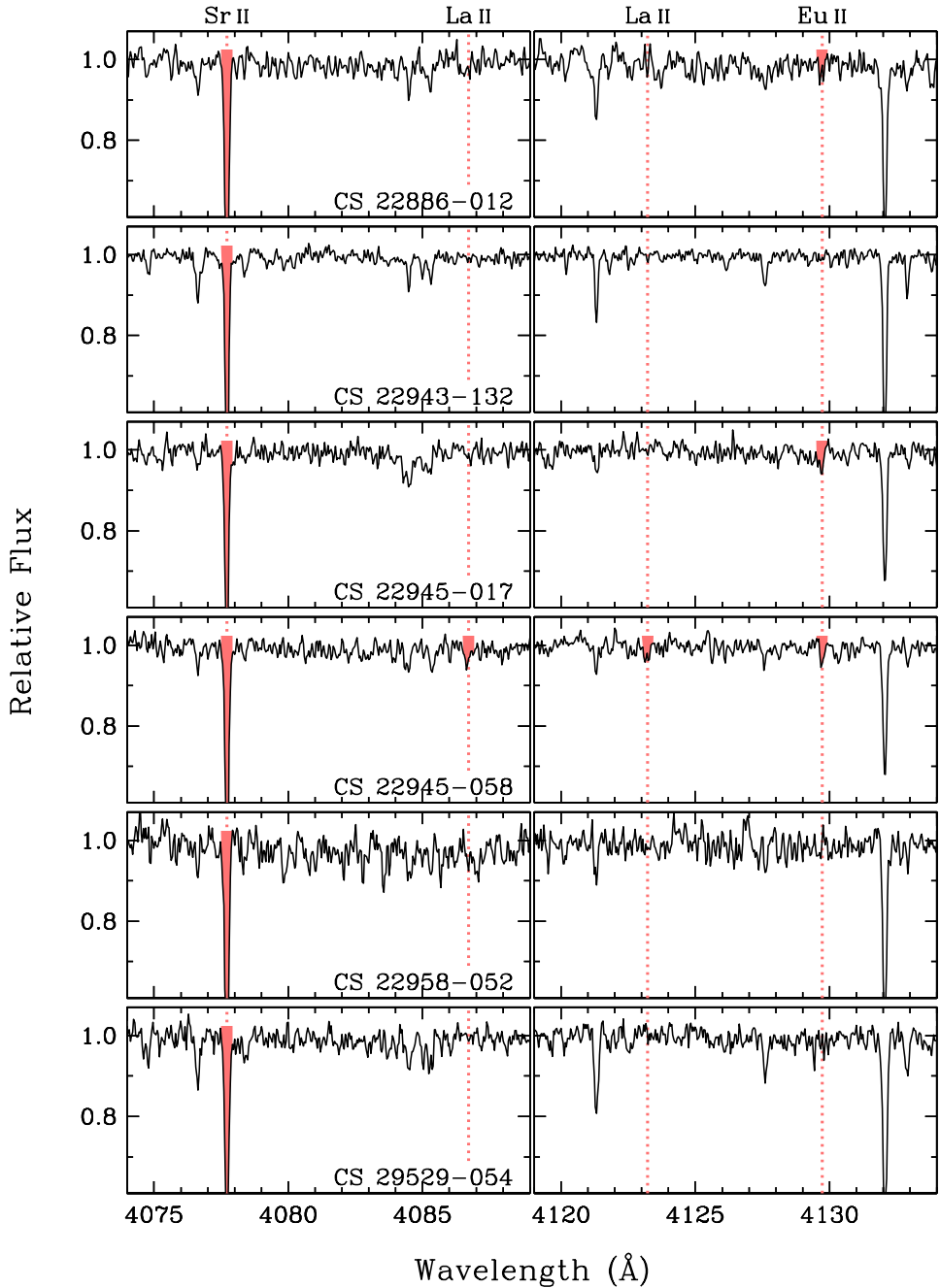}
\end{center}
\caption{
\label{sgbspecplot}
Spectra of the six subgiants.
Shaded regions indicate lines used in the abundance analysis
of \ncap\ elements.
Species are indicated at the top.
}
\end{figure*}

\begin{figure*}
\begin{center}
\includegraphics[angle=0,width=5.0in]{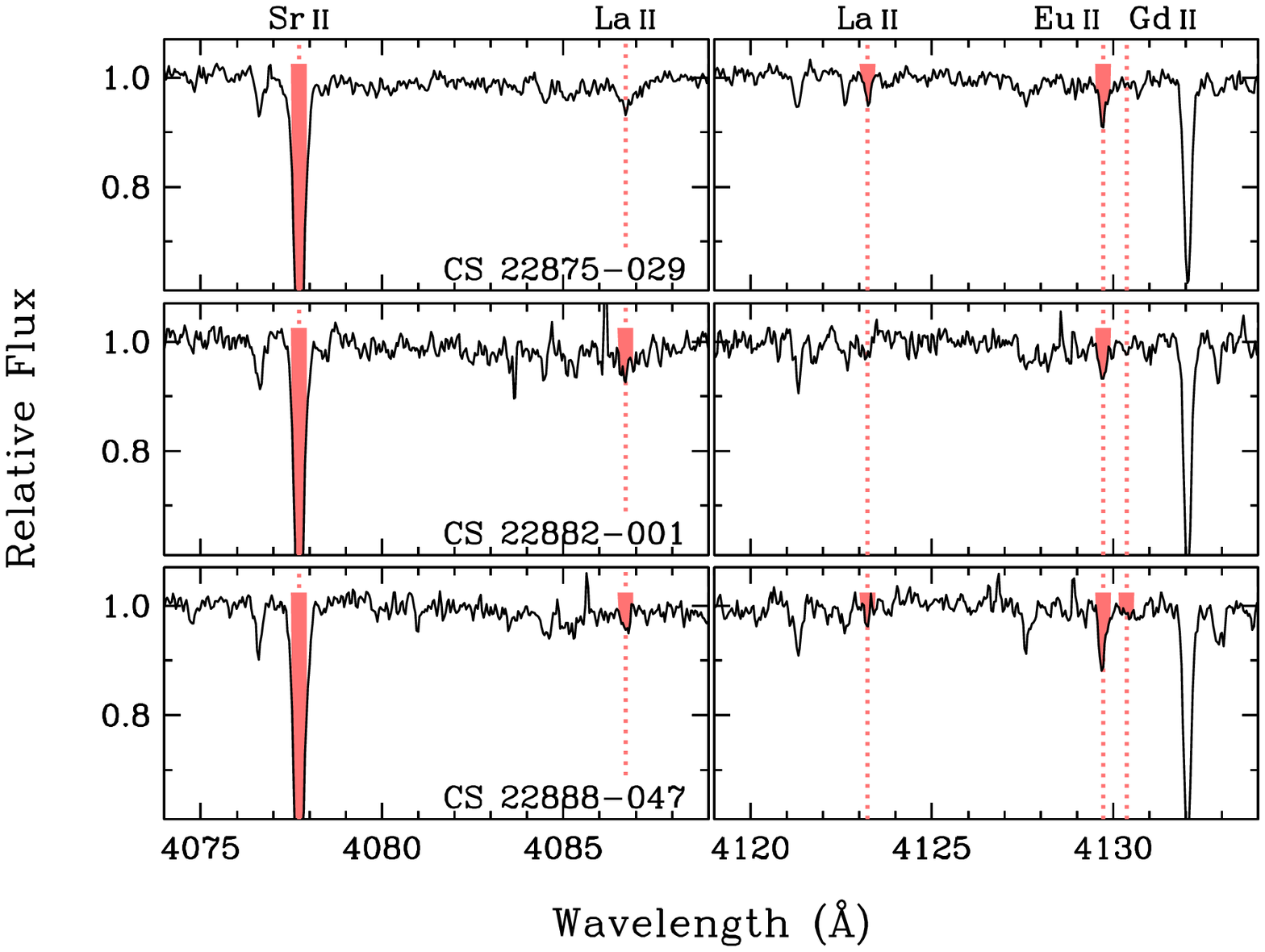}
\end{center}
\caption{
\label{rhbspecplot}
Spectra of the three stars on the RHB.~
Shaded regions indicate lines used in the abundance analysis
of \ncap\ elements.
Species are indicated at the top.
}
\end{figure*}

Portions of the spectra of the 13 
highly-\rpro-enhanced stars are shown in 
Figures~\ref{rgbspecplot}--\ref{rhbspecplot}.
The \ncap\ absorption lines are substantially stronger in the 
cool giants,
and there are many lines present.
This illustrates why highly-\rpro-enhanced stars 
have been preferentially identified among cool giants.
Absorption lines that are weak in cool giants
would be swamped by the continuous opacity in warmer stars,
rendering them undetectable.
None of these subgiants or RHB stars were selected for observation
based on the strength of their \ncap\ absorption lines.

Two other stars have [Eu/Fe] ratios in excess of the lower limit 
established to identify highly-\rpro-enhanced stars
(\mbox{CS~22956--102} and 
 \mbox{CS~29497--030}).
Their abundance patterns are clearly indicative
of enrichment by large amounts of material produced by
\spro\ nucleosynthesis.
These stars are marked with blue squares in Figure~\ref{euplot} 
and discussed separately in the Appendix.
We shall not consider them further here.

\section{Summary of Observations and Analysis Techniques}
\label{r14summary}

Table~\ref{obstab} presents a record of observations.
Observations were made with the Magellan Inamori Kyocera Echelle (MIKE)
spectrograph \citep{bernstein03}
on the 
6.5~m Walter Baade and Landon Clay Telescopes
(Magellan~I and II)
at Las Campanas Observatory.
These spectra were taken with the 0\farcs7\,$\times$\,5\farcs0 slit, 
yielding
a resolving power of $R \equiv \lambda/\Delta\lambda \sim$~41,000 in the blue 
and $R \sim$~35,000 in the red as measured from isolated ThAr lines
in the comparison lamp images.
The red and blue arms are split by a dichroic at $\approx$~4950~\AA.
This setup achieves complete wavelength coverage from 
3350--9150~\AA.
Signal to noise (S/N) estimates at several reference wavelengths
are listed in Table~\ref{sntab}.
The footnotes to Table~\ref{sntab} identify 
stars that were reimaged during the course
of the Las Campanas Objective-Prism Survey
\citep{beers92} or the
Hamburg-ESO Survey \citep{christlieb08}.

\begin{table*}
\begin{minipage}{\textwidth}
\caption{Log of Observations
\label{obstab}}
\begin{tabular}{lcccccc}
\hline
Star & 
Telescope/ &
Exposure &
Date &
UT at &
Heliocentric &
Heliocentric \\
 &
instrument &
length &
 &
mid-exposure &
Julian date &
radial velocity \\
 &
 &
 (s) &
 &
 & 
 &
(\kmsec) \\
\hline
\multicolumn{7}{c}{Previously-known $r$-process enhanced stars on the red giant branch} \\
\hline
CS 22183--031   & Magellan-Clay/MIKE      & 1400   & 2006 Aug 03 & 10:23 & 2453950.935   & $+$23.1       \\
CS 22892--052   & Magellan-Baade/MIKE     & 2400   & 2003 Jun 26 & 10:12 & 2452816.928   & $+$12.9       \\
                & Magellan-Clay/MIKE      & 2400   & 2004 Jun 29 & 09:11 & 2453185.886   & $+$13.0       \\
CS 22953--003   & Magellan-Clay/MIKE      & 2000   & 2006 Aug 03 & 09:46 & 2453950.910   & $+$208.3      \\
CS 31082--001   & Magellan-Clay/MIKE      & 1800   & 2003 Jul 19 & 10:27 & 2452839.937   & $+$138.9      \\
\hline
\multicolumn{7}{c}{New $r$-process enhanced stars on the subgiant branch or MSTO} \\
\hline
CS 22886--012   & Magellan-Clay/MIKE      & 3600   & 2008 Sep 12 & 02:59 & 2454721.630   & $-$56.8       \\
                & Magellan-Clay/MIKE      & 1800   & 2009 Sep 05 & 03:08 & 2455079.636   & $-$57.0       \\
CS 22943--132   & Magellan-Clay/MIKE      & 1400   & 2006 Aug 03 & 02:19 & 2453950.602   & $+$18.9       \\
CS 22945--017   & Magellan-Clay/MIKE      & 3400   & 2008 Sep 11 & 02:58 & 2454720.627   & $+$101.8      \\
CS 22945--058   & Magellan-Clay/MIKE      & 5400   & 2008 Sep 11 & 04:28 & 2454720.689   & $+$23.1       \\
                & Magellan-Clay/MIKE      & 1550   & 2009 Jul 26 & 10:29 & 2455038.940   & $+$23.7       \\
CS 22958--052   & Magellan-Clay/MIKE      & 3000   & 2004 Sep 26 & 09:12 & 2453274.886   & $+$88.9       \\
CS 29529--054   & Magellan-Baade/MIKE     & 3600   & 2003 Jan 17 & 03:27 & 2452656.643   & $+$113.3      \\
                & Magellan-Baade/MIKE     & 6000   & 2003 Jan 20 & 01:08 & 2452659.547   & $+$112.9      \\
                & Magellan-Clay/MIKE      & 1900   & 2009 Sep 05 & 09:59 & 2455079.917   & $+$113.4      \\
\hline
\multicolumn{7}{c}{New $r$-process enhanced stars on the red horizontal branch} \\
\hline
CS 22875--029   & Magellan-Baade/MIKE     & 3600   & 2003 Jun 12 & 09:59 & 2452802.918   & $+$73.1       \\
                & Magellan-Clay/MIKE      & 3600   & 2003 Oct 12 & 04:25 & 2452924.687   & $+$72.7       \\
CS 22882--001   & Magellan-Baade/MIKE     & 4800   & 2003 Jun 15 & 09:24 & 2452805.892   & $+$26.9       \\
                & Magellan-Clay/MIKE      & 6000   & 2003 Oct 09 & 06:05 & 2452921.758   & $+$26.1       \\
CS 22888--047   & Magellan-Baade/MIKE     & 4400   & 2003 Jun 14 & 09:00 & 2452804.876   & $-$158.5      \\
                & Magellan-Clay/MIKE      & 6000   & 2003 Oct 13 & 05:15 & 2452925.723   & $-$158.1      \\
\hline
\end{tabular}
\end{minipage}
\end{table*}

\begin{table*}
\begin{minipage}{\textwidth}
\caption{Total Exposure Times and S/N Ratios
\label{sntab}}
\begin{tabular}{lcccccc}
\hline
Star &
Total exp.\ &
No.\ &
S/N &
S/N &
S/N &
S/N \\
 &
time (s) &
obs.\ &
3950~\AA\ &
4550~\AA\ &
5200~\AA\ &
6750~\AA\ \\
\hline
\multicolumn{7}{c}{Previously-known $r$-process enhanced stars on the red giant branch} \\
\hline
CS 22183--031                    & 1400  & 1     & 75    & 105   & 75    & 105    \\
CS 22892--052\footnote{HE~2214$-$1654}   
                                 & 4800  & 2     & 150   & 225   & 175   & 270    \\
CS 22953--003                    & 2000  & 1     & 65    & 95    & 75    & 100    \\
CS 31082--001                    & 1800  & 1     & 125   & 145   & 230   & 350    \\
\hline
\multicolumn{7}{c}{New $r$-process enhanced stars on the subgiant branch} \\
\hline
CS 22886--012\footnote{CS~29512--015}
                                 & 5400  & 2     & 60    & 85    & 80    & 130    \\
CS 22943--132                    & 1400  & 1     & 75    & 105   & 75    & 95     \\ 
CS 22945--017                    & 3400  & 1     & 60    & 75    & 75    & 115    \\
CS 22945--058                    & 6950  & 2     & 60    & 85    & 80    & 125    \\
CS 22958--052                    & 3000  & 1     & 80    & 105   & 55    & 75     \\
CS 29529--054                    & 11500 & 3     & 60    & 100   & 110   & 155    \\
\hline
\multicolumn{7}{c}{New $r$-process enhanced stars on the red horizontal branch} \\
\hline
CS 22875--029                    & 7200  & 2     & 110   & 115   & 160   & 220    \\
CS 22882--001                    & 10800 & 2     & 85    & 95    & 120   & 170    \\
CS 22888--047                    & 10400 & 2     & 95    & 105   & 150   & 205    \\
\hline
\end{tabular}
\end{minipage}
\end{table*}

\citet{roederer14} used model atmospheres interpolated from the
grid of one-dimensional \textsc{marcs} models
\citep{gustafsson08}
and performed the analysis using a recent version of the
spectral line analysis code \textsc{moog}
(\citealt{sneden73}; see discussion in \citealt{sobeck11}).
Effective temperatures (\teff) and 
microturbulent velocities (\vt) were derived by requiring that 
abundances derived from Fe~\textsc{i} lines showed no trend with
the excitation potential 
and line strength.
For stars on the horizontal branch, 
\logg\ was derived by requiring that the Fe abundance derived
from Fe~\textsc{i} lines matched that derived from Fe~\textsc{ii} lines.
For the other stars,
surface gravities (\logg, in cgs units) were calculated from the 
relationship between \teff\ and \logg\ given by theoretical
isochrones in the Y$^{2}$ grid \citep{demarque04}
assuming an age of 12~$\pm$~1.5~Gyr.
The Fe abundance derived 
from Fe~\textsc{ii} lines was taken to represent
the overall metallicity, [M/H].
The derived model parameters and their statistical (internal)
uncertainties are presented in Table~\ref{atmtab}.
These values are identical to those presented in 
Table~7 of \citeauthor{roederer14}\ and are reproduced here for convenience.
\citeauthor{roederer14}\ estimated the systematic uncertainties
by comparing model parameters with those derived
in previous studies.
For red giants, subgiants, and stars on the horizontal branch,
these comparisons for the full sample yielded standard deviations of 
151, 211, and 156~K in \teff,
0.40, 0.34, and 0.42 in \logg,
0.41, 0.33, and 0.26~\kmsec\ in \vt, and
0.24, 0.22, and 0.16~dex in [Fe~\textsc{ii}/H].
We adopt these as the systematic uncertainties
in the model atmosphere parameters.

\begin{table*}
\begin{minipage}{\textwidth}
\caption{Atmospheric Parameters and Select Abundance Ratios Sorted by Decreasing [Eu/Fe] Ratios
\label{atmtab}}
\begin{tabular}{ccccccccc}
\hline
Star &
\teff\ &
\logg\ &
\vt\ &
[Fe/H]\footnote{As derived from Fe~\textsc{ii} lines} &
[Sr/Fe] &
[Ba/Fe] &
[Eu/Fe] &
[Yb/Fe] \\
 &
(K) &
 &
(\kmsec) &
 &
 &
 &
 &
  \\
\hline
\multicolumn{9}{c}{Previously-known $r$-process enhanced stars on the red giant branch} \\
\hline
 CS~31082--001  & 4650 (35) & 1.05 (0.14) & 1.55 (0.06) & $-$3.03 (0.07) & $+$0.67 (0.12) & $+$0.92 (0.17) & $+$1.37 (0.16) & $+$1.38 (0.27) \\
 CS~22892--052  & 4690 (37) & 1.15 (0.15) & 1.50 (0.06) & $-$3.16 (0.07) & $+$0.68 (0.17) & $+$0.94 (0.16) & $+$1.35 (0.22) & $+$1.39 (0.27) \\
 CS~22183--031  & 4850 (36) & 1.60 (0.14) & 1.55 (0.06) & $-$3.50 (0.07) & $-$0.22 (0.23) & $+$0.13 (0.15) & $+$0.84 (0.16) & $+$0.68 (0.18) \\
 CS~22953--003  & 4860 (37) & 1.65 (0.17) & 1.45 (0.06) & $-$3.00 (0.07) & $+$0.32 (0.24) & $+$0.29 (0.15) & $+$0.79 (0.15) & $+$0.63 (0.20) \\
\hline
\multicolumn{9}{c}{New $r$-process enhanced stars on the subgiant branch} \\
\hline
 CS~22945--017  & 6080 (54) & 3.70 (0.19) & 1.25 (0.06) & $-$2.73 (0.08) & $+$0.39 (0.24) & $+$0.49 (0.14) & $+$1.13 (0.19) & $+$1.02 (0.18) \\
 CS~22945--058  & 5990 (45) & 3.65 (0.17) & 1.55 (0.06) & $-$2.71 (0.08) & $+$0.34 (0.24) & $+$0.28 (0.14) & $+$1.13 (0.19) & $+$0.83 (0.20) \\
 CS~22958--052  & 6090 (46) & 3.75 (0.18) & 1.95 (0.06) & $-$2.42 (0.07) & $+$0.13 (0.24) & $+$0.00 (0.15) & $+$1.00 (0.18) & $+$0.93 (0.17) \\
 CS~29529--054  & 5710 (40) & 3.55 (0.18) & 1.65 (0.06) & $-$2.75 (0.07) & $-$0.10 (0.24) & $-$0.02 (0.15) & $+$0.90 (0.21) & $+$0.86 (0.24) \\
 CS~22943--132  & 5850 (41) & 3.60 (0.16) & 1.40 (0.06) & $-$2.67 (0.08) & $+$0.27 (0.24) & $-$0.05 (0.14) & $+$0.86 (0.22) & $+$0.37 (0.32) \\
 CS 22886--012  & 5650 (42) & 3.50 (0.15) & 1.45 (0.06) & $-$2.61 (0.07) & $+$0.31 (0.23) & $+$0.14 (0.14) & $+$0.85 (0.17) & $+$0.45 (0.22) \\
\hline
\multicolumn{9}{c}{New $r$-process enhanced stars on the horizontal branch} \\
\hline
 CS~22875--029  & 5990 (44) & 1.85 (0.27) & 2.80 (0.06) & $-$2.69 (0.06) & $+$0.85 (0.22) & $+$0.30 (0.15) & $+$0.92 (0.16) & $+$0.86 (0.19) \\
 CS~22888--047  & 5950 (46) & 1.90 (0.27) & 3.00 (0.06) & $-$2.54 (0.06) & $+$0.60 (0.22) & $+$0.04 (0.16) & $+$0.86 (0.16) & $+$0.67 (0.18) \\
 CS~22882--001  & 5930 (52) & 1.90 (0.32) & 3.00 (0.06) & $-$2.62 (0.06) & $+$0.16 (0.25) & $+$0.06 (0.16) & $+$0.81 (0.16) & $+$0.56 (0.19) \\
\hline
\multicolumn{9}{c}{Mean ratios for all 13 stars} \\
\hline
  \ldots &  \ldots &  \ldots &  \ldots & $-$2.80 (0.08) & $+$0.34 (0.09) & $+$0.27 (0.09) & $+$0.99 (0.06) & $+$0.82 (0.09) \\
\hline
\end{tabular}
\end{minipage}
\end{table*}

Table~8 of \citet{roederer14} lists the 
atomic data for each transition studied.
Spectrum synthesis was performed for lines broadened by
hyperfine splitting or in cases where
a significant isotope shift may be present.
For unblended lines, \citeauthor{roederer14}\ used \textsc{moog} to compute
theoretical equivalent widths, which were then forced to match
measured equivalent widths by adjusting the abundance.
\citeauthor{roederer14}\ derived
3$\sigma$ upper limits 
when a line was not detected.
Table~11 of \citeauthor{roederer14}\ lists the abundances
derived from each line in each star.
That study also adopted corrections
to account for departures from local thermodynamic equilibrium (LTE) 
in the line formation regions for
Li~\textsc{i} \citep{lind09},
O~\textsc{i} \citep{fabbian09}, 
Na~\textsc{i} \citep{lind11}, and 
K~\textsc{i} \citep{takeda02}.
Weighted mean abundances and uncertainties were computed using the
formalism presented in \citet{mcwilliam95b}.
We do not repeat the full set of abundances here, but a few key
element ratios are reproduced in
Table~\ref{atmtab} for convenience.

\section{Comparison with Previous Work}
\label{compare}

Table~\ref{comparetab} compares the derived stellar parameters
and several abundance ratios
with those found by previous investigators
for the four red giant and three RHB stars in our sample.
Most previous studies of red giants calculated \teff\ 
using color-\teff\ relations, frequently leading to
warmer \teff\ and higher metallicity than that found by \citet{roederer14}.
For the four red giant stars listed in Table~\ref{comparetab}, 
our metallicities are lower than those found by previous studies
by 0.19~dex ($\sigma =$~0.13).
For the RHB stars, our metallicities are lower than those found
by previous studies by 0.09~dex ($\sigma =$~0.10).
No comparisons are available for the subgiants,
but \citeauthor{roederer14}\ found that the derived metallicities
of 40~subgiants were lower than those found by previous studies
by only 0.04~dex ($\sigma =$~0.18).

The [Eu/Fe] ratio is often adopted as a proxy for the overall
level of \rpro\ enhancement of a star, and our [Eu/Fe] ratios
are lower than found by previous studies by
0.22~dex ($\sigma =$~0.10) for the red giants and lower by
0.09~dex ($\sigma =$~0.08) for the stars on the RHB.
The reasons for these differences are investigated in detail
in Section~9.4 of \citeauthor{roederer14}~ 
In summary, this offset can be explained by the combined effects of
different model atmosphere parameters,
updates to the analysis code, lines available for analysis,
and quality of the observed spectra.

\begin{table*}
\begin{minipage}{\textwidth}
\caption{Comparison of Derived Model Parameters
and Abundances with Previous Work
\label{comparetab}}
\begin{tabular}{cccccccccc}
\hline
Star &
\teff\ &
\logg\ &
\vt\ &
[Fe/H]\footnote{As derived from Fe~\textsc{ii} lines, if specified} &
[Sr/Fe] &
[Ba/Fe] &
[Eu/Fe] &
[Yb/Fe] &
Ref.\ \\
 &
(K) &
 & 
(\kmsec) &
 & 
 & 
 & 
 & 
 & 
 \\
\hline
 CS~22183--031  & 4850 & 1.60 & 1.55 & $-$3.50 & $-$0.22 & $+$0.13 & $+$0.84 & $+$0.68 & 1  \\
 (RG)           & 5270 & 2.80 & 1.20 & $-$2.93 & $+$0.10 & $+$0.38 & $+$1.16 &  \ldots & 2  \\
\hline
 CS~22892--052  & 4690 & 1.15 & 1.50 & $-$3.16 & $+$0.68 & $+$0.94 & $+$1.35 & $+$1.39 & 1  \\
 (RG)           & 4790 & 1.60 & 1.80 & $-$2.92 & $+$0.44 & $+$0.92 & $+$1.51 & $+$1.29 & 2  \\
                & 4760 & 1.30 & 2.29 & $-$3.03 & $+$0.68 & $+$0.93 & $+$1.48 &  \ldots & 3, 4 \\
                & 4850 & 1.60 & 1.90 & $-$3.02 & $+$0.53 & $+$1.01 & $+$1.49 &  \ldots & 5, 6 \\
                & 4800 & 1.50 & 1.95 & $-$3.12 & $+$0.58 & $+$0.92 & $+$1.65 & $+$1.63 & 7, 8 \\
                & 4884 & 1.81 & 1.67 & $-$2.95 & $+$0.61 & $+$1.19 & $+$1.54 &  \ldots & 9 \\
                & 4725 & 1.00 & 2.00 & $-$3.16 & $+$0.52 & $+$0.89 & $+$1.56 &  \ldots & 10 \\
                & 4760 & 1.30 & 2.30 & $-$3.03 & $+$0.63 & $+$0.89 & $+$1.63 & $+$1.15 & 11 \\
\hline
 CS~22953--003  & 4860 & 1.65 & 1.45 & $-$3.00 & $+$0.32 & $+$0.29 & $+$0.79 & $+$0.63 & 1  \\
 (RG)           & 4960 & 1.70 & 1.95 & $-$2.83 & $+$0.31 & $+$0.00 & $+$0.72 &  \ldots & 3, 4 \\
                & 5100 & 2.30 & 1.70 & $-$2.82 & $+$0.22 & $+$0.49 & $+$1.05 & $+$1.02 & 5, 6 \\
\hline
 CS~31082--001  & 4650 & 1.05 & 1.55 & $-$3.03 & $+$0.67 & $+$0.92 & $+$1.37 & $+$1.38 & 1  \\
 (RG)           & 4825 & 1.50 & 1.80 & $-$2.92 & $+$0.65 & $+$1.17 & $+$1.63 &  \ldots & 12 \\
                & 4790 & 1.80 & 1.90 & $-$2.81 & $+$0.47 & $+$1.02 & $+$1.67 & $+$1.54 & 2  \\
                & 4825 & 1.50 & 1.80 & $-$2.92 & $+$0.73 & $+$1.16 & $+$1.69 & $+$1.66 & 13 \\
                & 4825 & 1.50 & 1.80 & $-$2.92 &  \ldots &  \ldots & $+$1.68 & $+$1.57 & 8  \\
                & 4922 & 1.90 & 1.88 & $-$2.78 & $+$0.53 & $+$1.18 & $+$1.66 &  \ldots & 9  \\
                & 4925 & 1.51 & 1.40 & $-$2.81 & $+$0.66 & $+$1.43 & $+$1.53 &  \ldots & 14 \\
\hline
 CS~22875--029  & 5990 & 1.85 & 2.80 & $-$2.69 & $+$0.85 & $+$0.30 & $+$0.92 & $+$0.86 & 1  \\
 (RHB)          & 6000 & 2.35 & 3.05 & $-$2.63 & $+$0.83 & $+$0.48 & $+$1.10 &  \ldots & 15 \\
                & 6000 & 2.05 & 3.00 & $-$2.66 & $+$0.86 & $+$0.44 & $+$0.91 &  \ldots & 16 \\
\hline
 CS~22882--001  & 5930 & 1.90 & 3.00 & $-$2.62 & $+$0.16 & $+$0.06 & $+$0.81 & $+$0.56 & 1  \\
 (RHB)          & 5950 & 2.50 & 3.20 & $-$2.45 & $+$0.28 & $+$0.20 & $+$0.92 &  \ldots & 15 \\
                & 5950 & 2.00 & 3.05 & $-$2.54 & $+$0.22 & $+$0.16 & $+$0.84 &  \ldots & 16 \\
\hline
 CS~22888--047  & 5950 & 1.90 & 3.00 & $-$2.54 & $+$0.60 & $+$0.04 & $+$0.86 & $+$0.67 & 1  \\
 (RHB)          & 6000 & 2.50 & 3.00 & $-$2.30 & $+$0.63 & $+$0.29 & $+$1.03 &  \ldots & 15 \\
                & 5850 & 1.70 & 3.20 & $-$2.57 & $+$0.31 & $+$0.23 & $+$0.93 &  \ldots & 16 \\
\hline
\multicolumn{10}{c}{Average differences for the four red giant stars} \\
\hline
Mean                & $-$169 & $-$0.46 & $-$0.33 & $-$0.19 & $+$0.06 & $-$0.11 & $-$0.22 & $-$0.13 & \\
Standard deviation  &    105 &    0.35 &    0.31 &    0.13 &    0.13 &    0.19 &    0.10 &    0.22 & \\
No.\ comparisons    & 14     & 14      & 14      & 14      & 15      & 15      & 16      &  7      & \\
\hline
\multicolumn{10}{c}{Average differences for the three stars on the red horizontal branch} \\
\hline
Mean                &   $-$2 & $-$0.30 & $-$0.15 & $-$0.09 & $+$0.02 & $-$0.05 & $-$0.09 &  \ldots & \\
Standard deviation  &     52 &    0.32 &    0.10 &    0.10 &    0.14 &    0.15 &    0.08 &  \ldots & \\
No. comparisons     &  6     &  6      &  6      &  6      &  6      &  6      &  6      &  \ldots & \\
\hline
\end{tabular}
\\
Studies that adopted model parameters without change from an earlier study have been excluded
from computing the average deviations of \teff, \logg, \vt, and [Fe/H]. \\
References:\
(1) This study;
(2) \citealt{honda04b};
(3) \citealt{mcwilliam95b};
(4) \citealt{mcwilliam98};
(5) \citealt{cayrel04};
(6) \citealt{francois07};
(7) \citealt{sneden03};
(8) \citealt{sneden09};
(9) \citealt{barklem05};
(10) \citealt{sneden94};
(11) \citealt{sneden96};
(12) \citealt{hill02};
(13) \citealt{siqueiramello13};
(14) \citealt{hansen12};
(15) \citealt{preston06};
(16) \citealt{for10}
\end{minipage}
\end{table*}

\section{Radial Velocity Measurements}
\label{rv}

\citet{roederer14} measured radial velocities
by cross-correlating the spectral
order containing the Mg~\textsc{i}~b lines
against metal-poor template standards.
Heliocentric corrections 
were computed using the \textsc{iraf} \textit{rvcorrect} task.
Typical uncertainties are
$\approx$~0.6--0.8~\kmsec\
per observation.
Table~\ref{obstab} lists the
Heliocentric velocity measurements for each observation.

Repeat observations of
\hba,
\hbb,
\sga,
\hbc,
\rgb,
\sgd,
and
\sgf\
show no evidence of radial velocity variations in our data.
\citet{preston01} reported a 
possible detection of binarity for \rgb, but
that preliminary result was based on a
tentative phasing of measurements with
semi-amplitude of only 1.0~\kmsec.
Subsequent velocity monitoring by our group and others
\citep{mcwilliam95a,honda04a,bonifacio09,hansen11}
finds no compelling evidence for
velocity variations of \rgb.
Comparisons with prior studies of
\rgc\
(\citeauthor{mcwilliam95a,bonifacio09})
and
\rgd\
(\citeauthor{honda04a,bonifacio09,hansen11})
also show no evidence of radial velocity variations.
Our measured velocity of \rga\ is 
11~\kmsec\ different from two measurements made by
\citet{honda04a}.
We are aware of only our single-epoch radial velocity measurements of
\sgb,
\sgc, and
\sge.

Of the 10~stars in our sample with observations at multiple epochs
from all sources,
only \rga\ shows compelling evidence for velocity variations.
This frequency, 10~per cent, 
is in reasonable agreement with the 18~per cent binary frequency (3 of 17~stars)
of other highly-\rpro-enhanced stars
reported by \citet{hansen11} based on long-term velocity monitoring.
Our binary frequency 
may be underestimated since we have velocity measurements
at only two or three epochs for most stars,
thus long-period or low-amplitude binaries may
evade our search.
In any case, our data support the conclusions of \citeauthor{hansen11}\
that there is no evidence to suggest that all
highly-\rpro-enhanced stars are members of binary or
multiple star systems.

\section{The $r$-process Abundance Patterns}
\label{patterns}

\begin{figure*}
\begin{center}
\includegraphics[angle=0,width=2.9in]{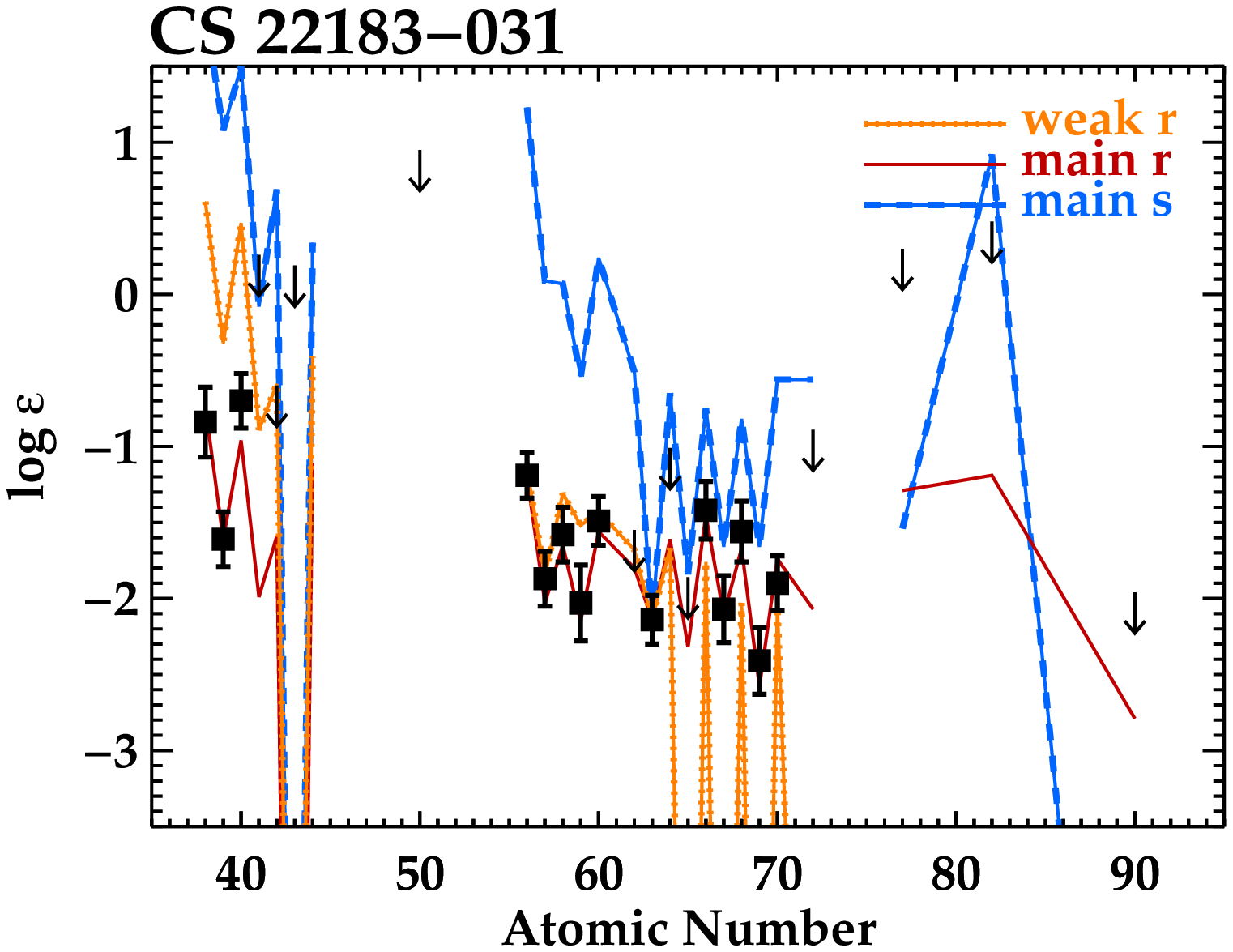}
\hspace*{0.3in}
\includegraphics[angle=0,width=2.9in]{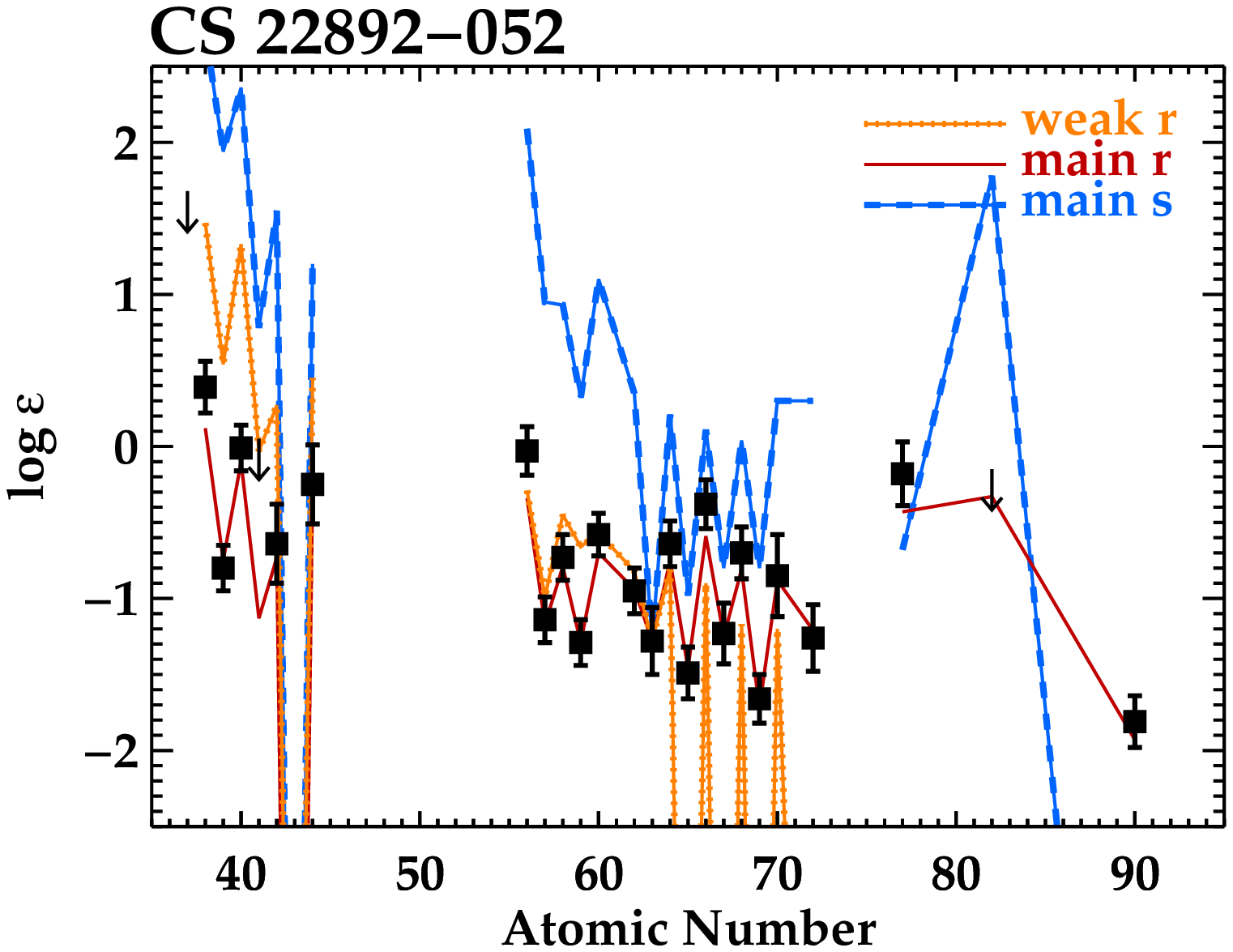} \\
\vspace*{0.3in}
\includegraphics[angle=0,width=2.9in]{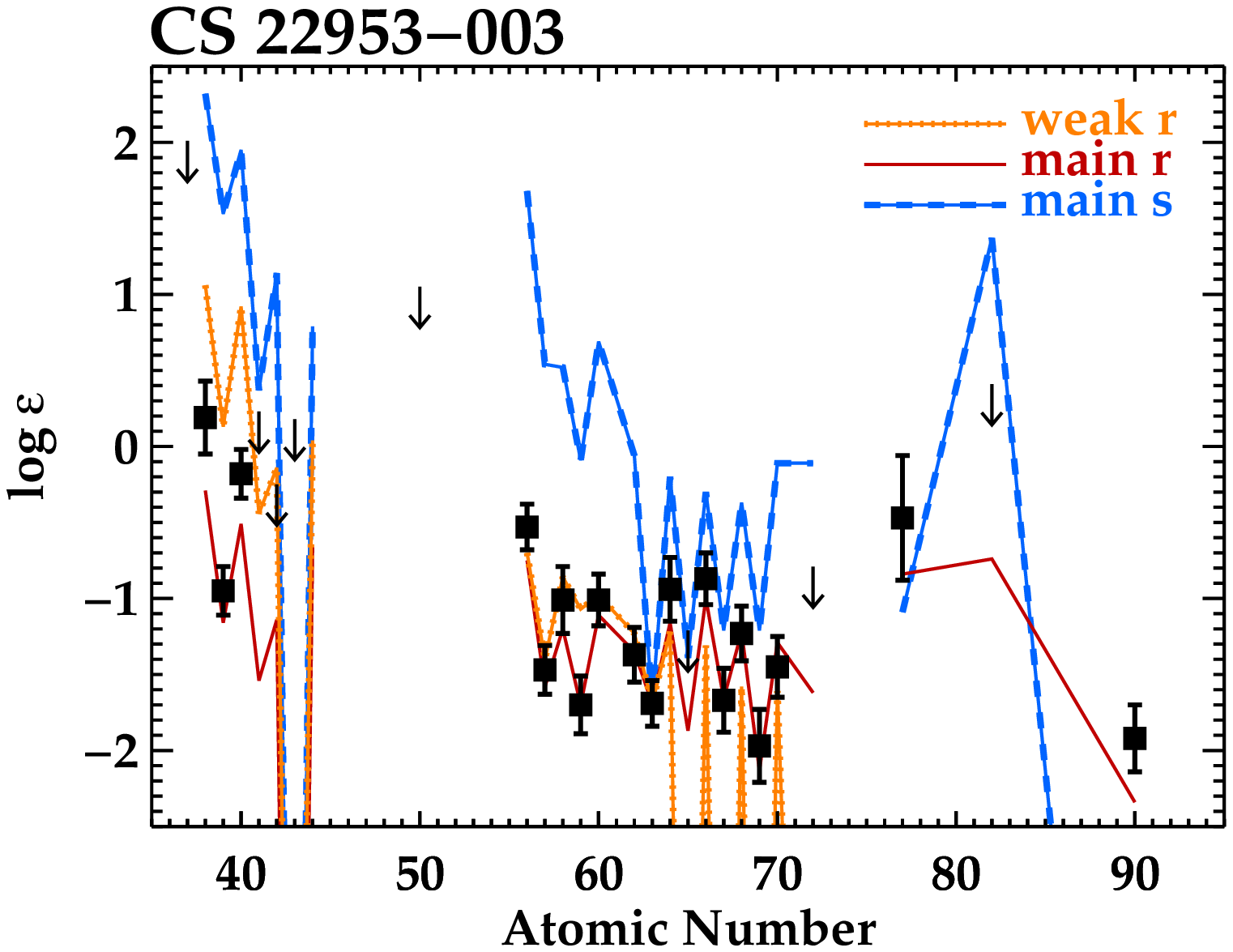}
\hspace*{0.3in}
\includegraphics[angle=0,width=2.9in]{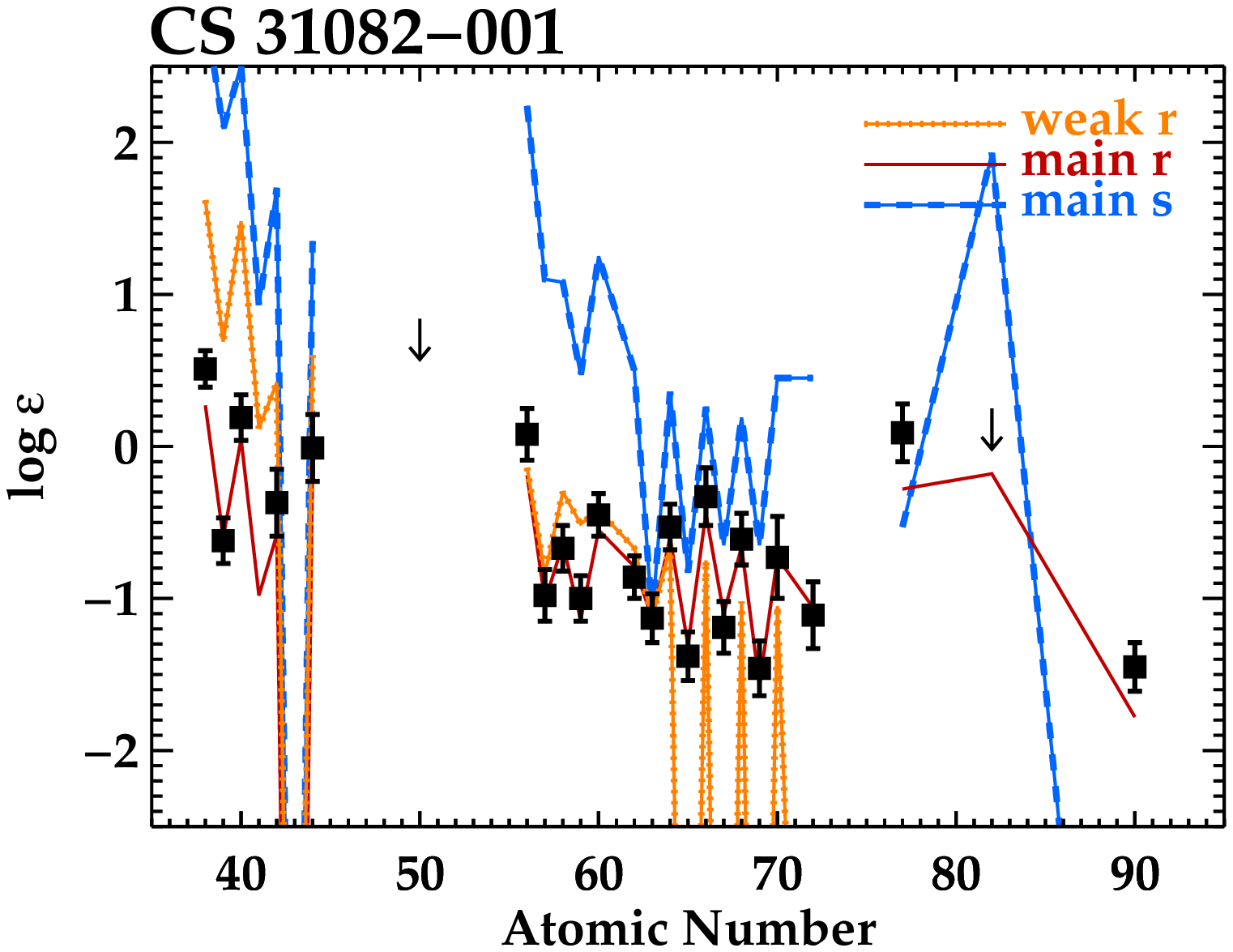} \\
\end{center}
\caption{
\label{rprorgfig}
The heavy element distributions in the four red giants:\
\mbox{CS~22183--031}, \mbox{CS~22892--052},
\mbox{CS~22953--003}, and \mbox{CS~31082--001}.
Filled squares mark detections, and arrows mark 3$\sigma$ upper limits
derived from non-detections.
The studded orange line marks the scaled heavy element
distribution found in the metal-poor giant \mbox{HD~122563}
\citep{honda06,roederer12},
frequently referred to as the weak component of the \rpro.
The solid red line marks the scaled heavy element
distribution found in \mbox{CS~22892--052},
the main component of the \rpro,
as derived previously by
\citet{sneden03,sneden09} and \citet{roederer09}.
The long-dashed blue line marks the scaled heavy element
distribution predicted by the main and strong components of the \spro.
Each of the three curves has been renormalized
to the Eu abundance.
}
\end{figure*}

\begin{figure*}
\begin{center}
\includegraphics[angle=0,width=2.9in]{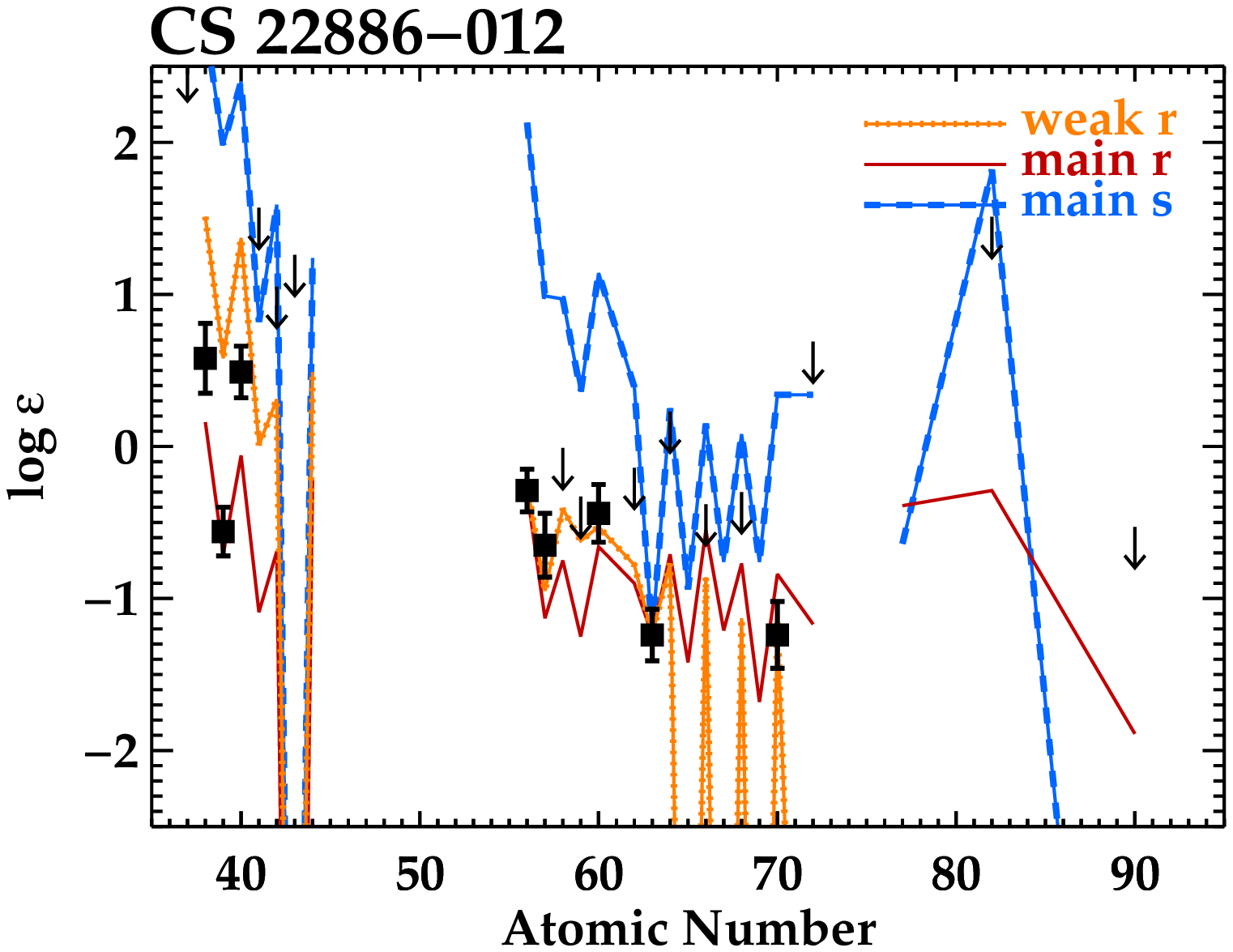}
\hspace*{0.3in}
\includegraphics[angle=0,width=2.9in]{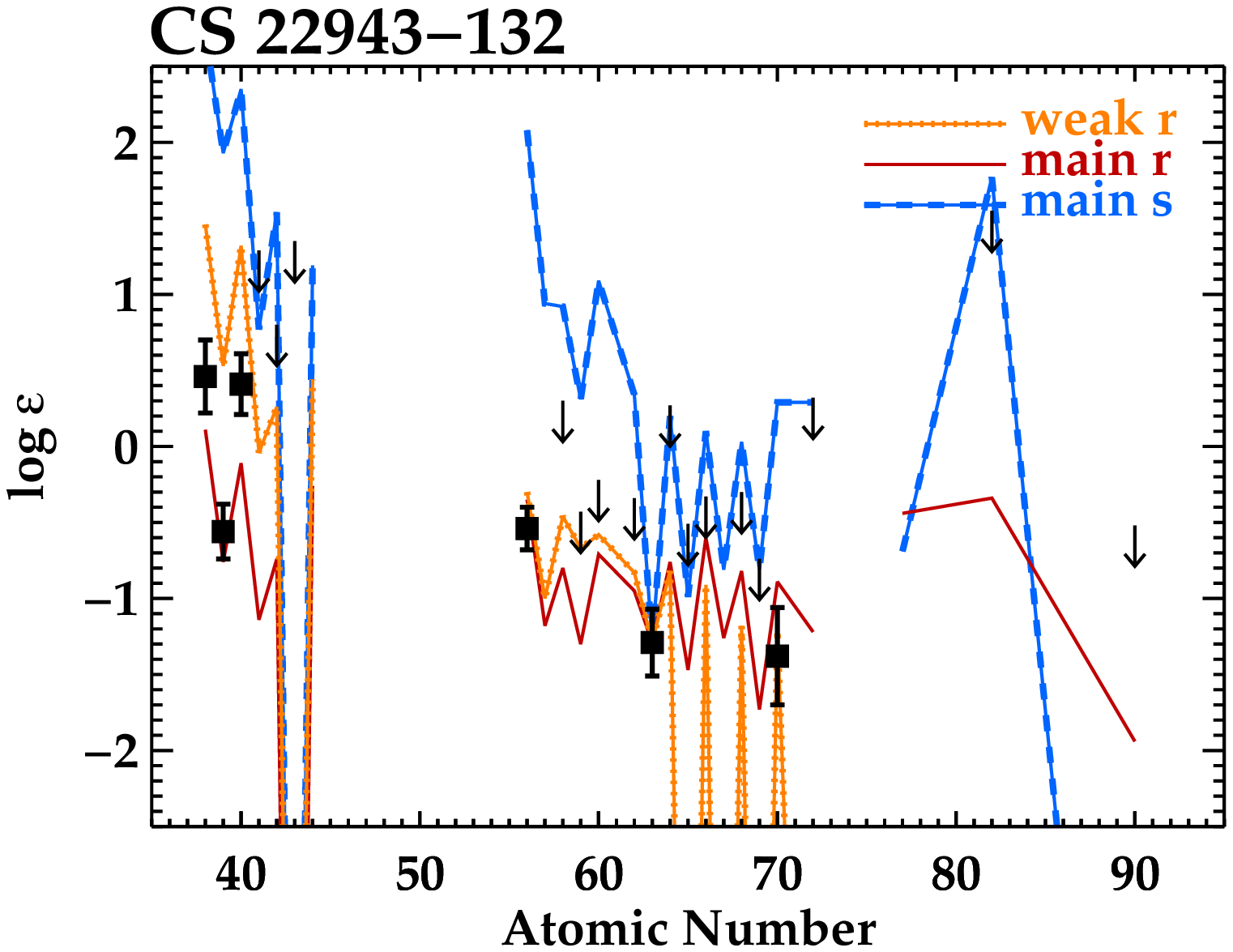} \\
\vspace*{0.3in}
\includegraphics[angle=0,width=2.9in]{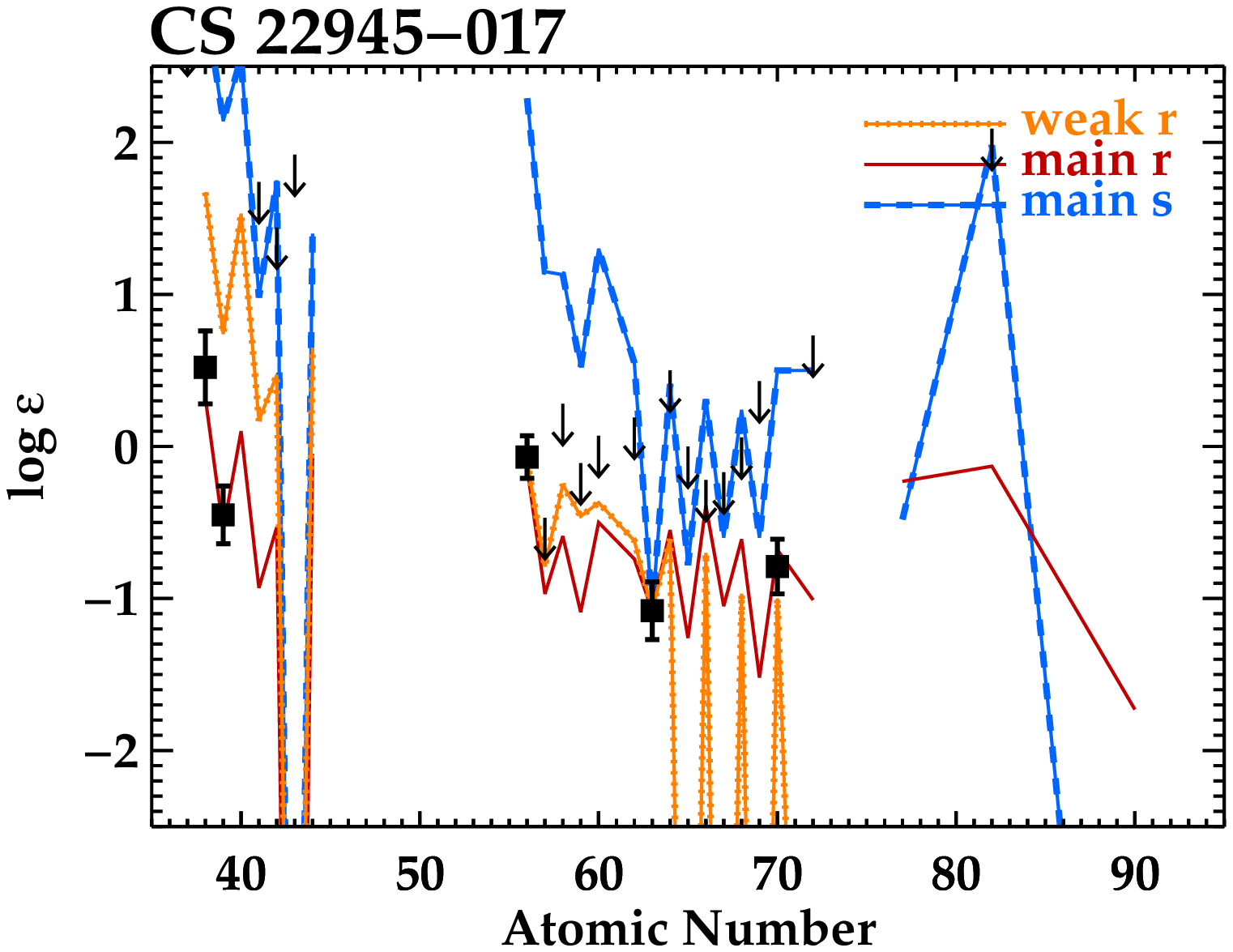}
\hspace*{0.3in}
\includegraphics[angle=0,width=2.9in]{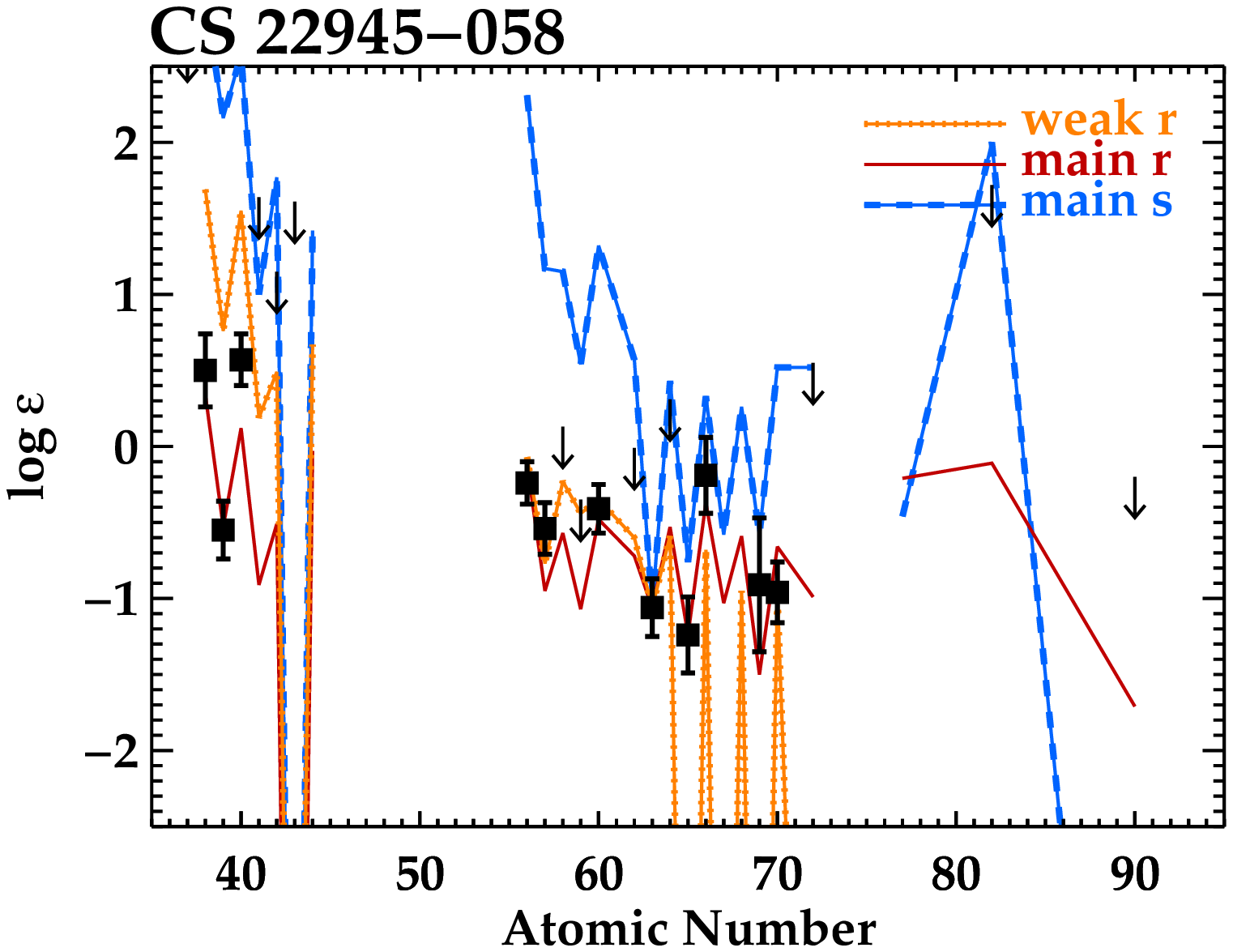} \\
\vspace*{0.3in}
\includegraphics[angle=0,width=2.9in]{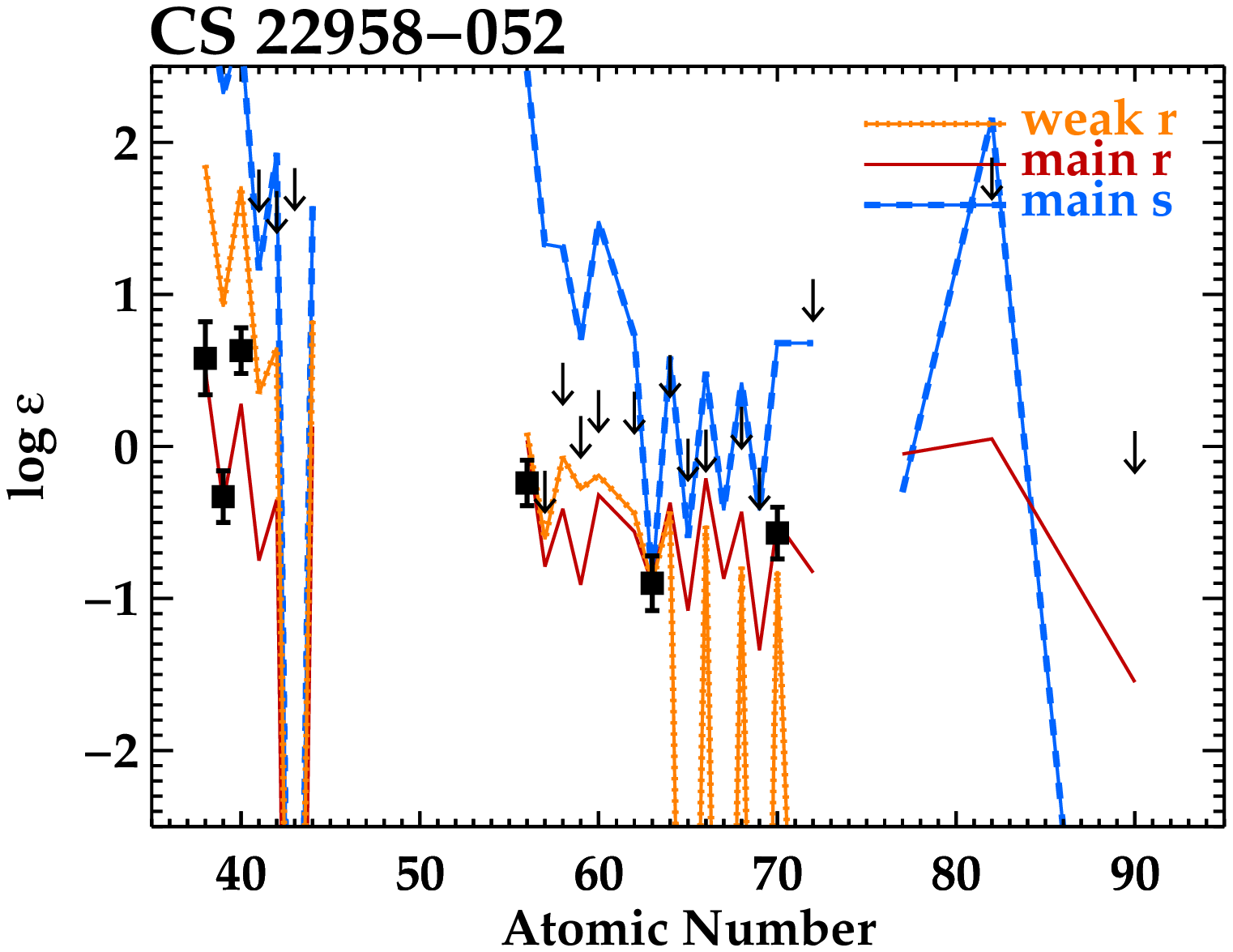}
\hspace*{0.3in}
\includegraphics[angle=0,width=2.9in]{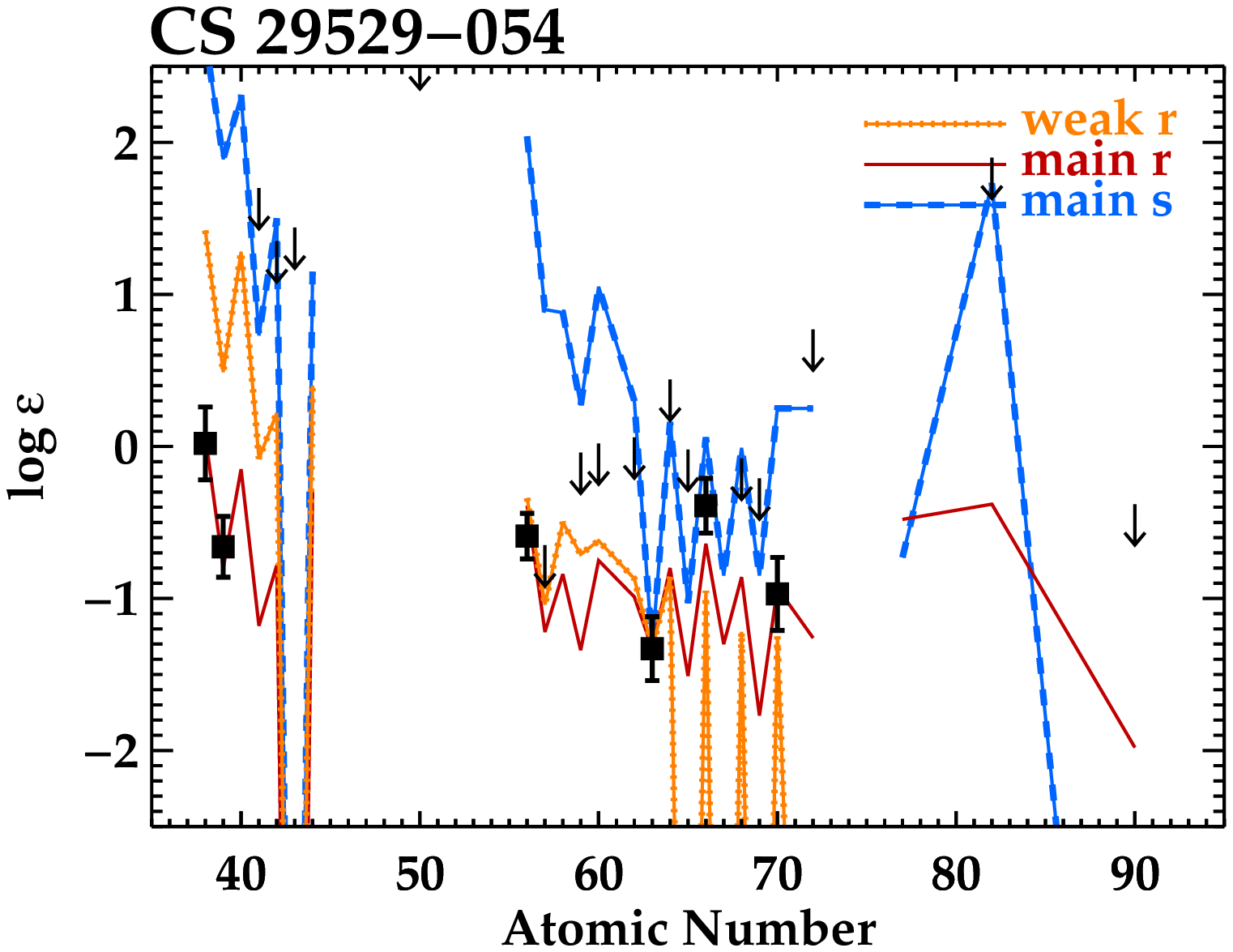} \\
\end{center}
\caption{
\label{rprosgfig}
The heavy element distributions in the six subgiants:\
\mbox{CS~22886--012}, \mbox{CS~22943--132}, \mbox{CS~22945--017}, 
\mbox{CS~22945--058}, \mbox{CS~22958--052}, and \mbox{CS~29529--054}.
Symbols are the same as in Figure~\ref{rprorgfig}.
}
\end{figure*}

\begin{figure}
\begin{center}
\includegraphics[angle=0,width=2.9in]{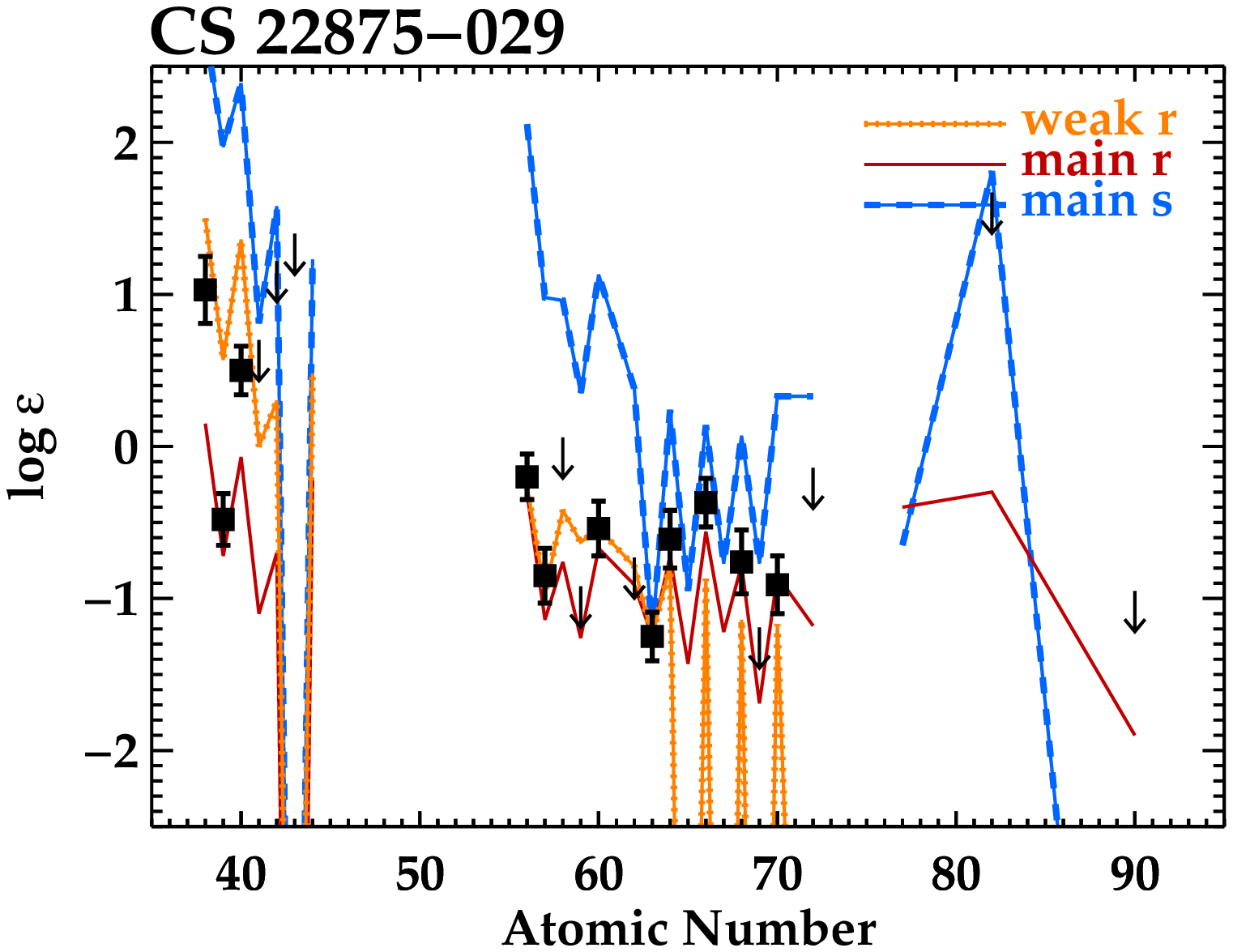} \\
\vspace*{0.3in}
\includegraphics[angle=0,width=2.9in]{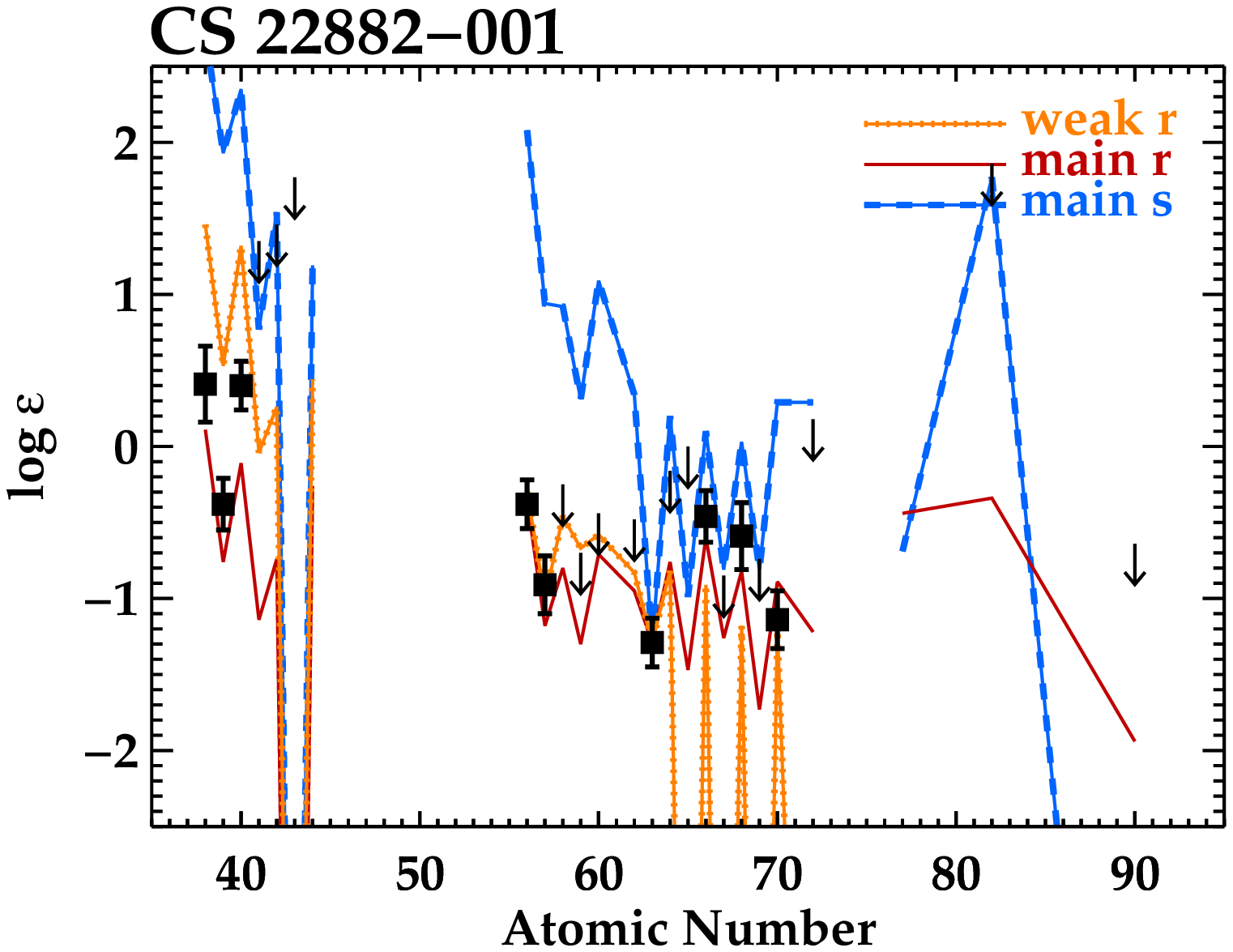} \\
\vspace*{0.3in}
\includegraphics[angle=0,width=2.9in]{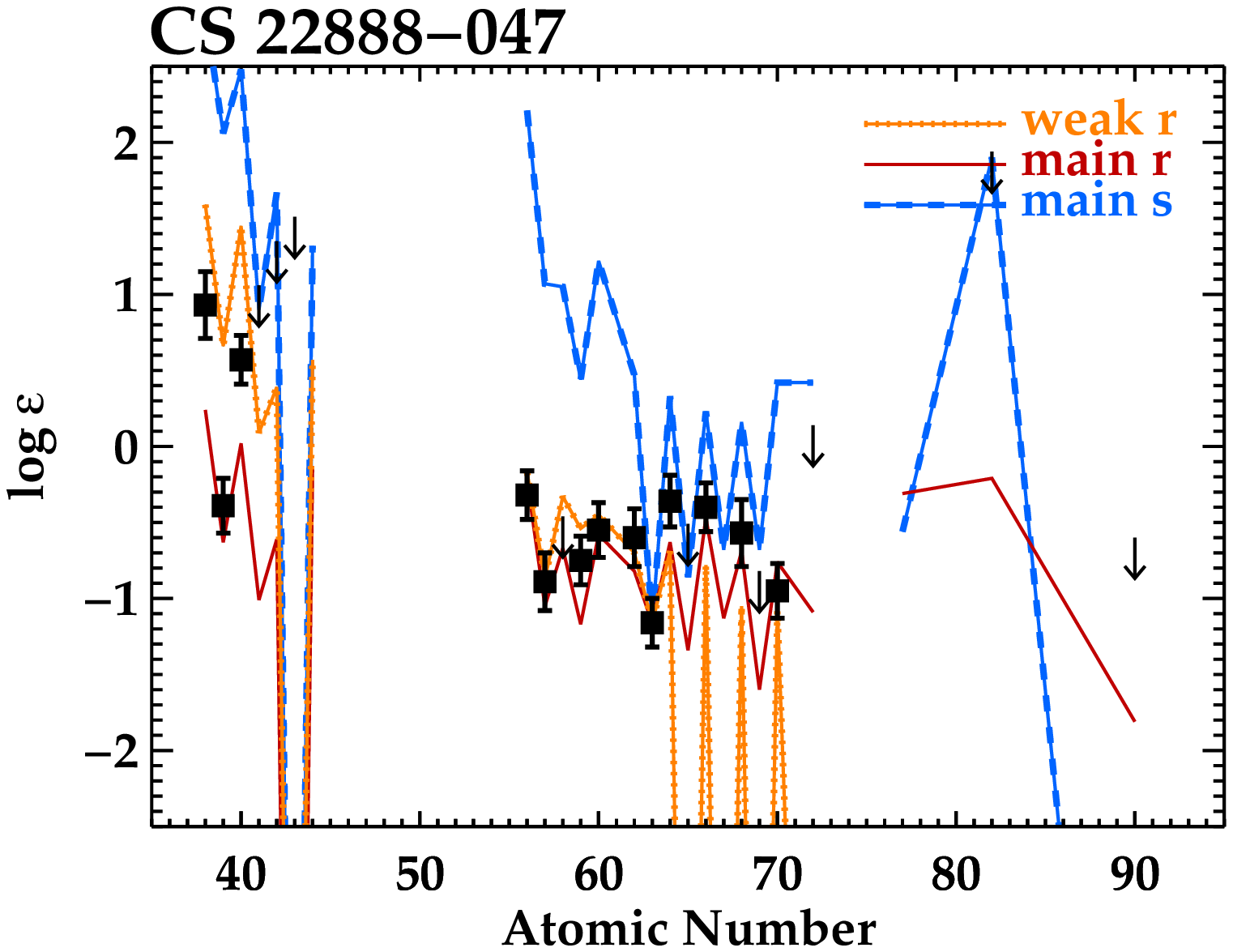} \\
\end{center}
\caption{
\label{rprohbfig}
The heavy element distributions in the three stars on the horizontal branch:\
\mbox{CS~22875--029}, \mbox{CS~22882--001}, and \mbox{CS~22888--047}.
Symbols are the same as in Figure~\ref{rprorgfig}.
}
\end{figure}

The heavy element abundance patterns are illustrated in 
Figures~\ref{rprorgfig}, \ref{rprosgfig}, and \ref{rprohbfig}.
Three standard abundance templates are shown for comparison.
One pattern traces the heavy element abundances
in \rgb\ as derived by \citet{sneden03,sneden09} and \citet{roederer09}
from higher quality data.
It is reassuring that our derived abundance pattern
for \rgb\ matches this comparison set so closely.
The small deviations for Sr and Ba can be
explained by the different
microturbulent velocities derived by us (1.50~\kmsec)
and \citeauthor{sneden03}\ (1.95~\kmsec),
which cause us to derive larger abundances from
Sr~\textsc{ii} and Ba~\textsc{ii} lines approaching saturation.
Another pattern traces the heavy element abundances
in the metal-poor halo star \mbox{HD~122563},
which has normal abundances of the lighter 
\ncap\ elements and a deficiency
of the heaviest \ncap\ elements.
The pattern found in this star may be
considered representative of the so-called
weak component of the \rpro.
The third pattern traces one outcome of \spro\ nucleosynthesis
\citep{sneden08,bisterzo11}.
In each figure, these patterns are rescaled to match the 
stellar Eu abundance.

The largest numbers of elements are detected in the
red giants, shown in Figure~\ref{rprorgfig}.
Nearly all of the rare earth elements 
(57~$\leq Z \leq$~71,
plus neighboring elements Ba and Hf)
are detected in each case.
Ir, a member of the third \rpro\ peak, and 
the actinide Th are also detected in three of these stars.
Among the lighter elements, Sr, Y, and Zr
are always detectable, and
Mo and Ru are detected in two of the stars.
Fewer elements are detectable in the subgiants
(Figure~\ref{rprosgfig}) or field RHB stars
(Figure~\ref{rprohbfig}).
Yet in all of these stars
the rare earth elements and those beyond 
match the pattern in \rgb\ established by previous studies.

Only when comparing the ratios of the lighter \ncap\ elements
with the rare earth elements do significant differences emerge.
This is revealed in Figures~\ref{rprorgfig}--\ref{rprohbfig} 
by the Sr, Y, and Zr
abundances that are sometimes enhanced relative to the
scaled main component of the \rpro\ (solid red line).
Numerous studies have pointed out this 
characteristic in other samples
of \rpro-enhanced field stars, 
among them \citet{mcwilliam98},
\citet{johnson02}, \citet{aoki05}, \citet{hansen12}, and
\citet{siqueiramello14}.
Curiously,
one \rpro-enhanced metal-poor star analyzed by \citet{roederer14c} 
hints that the so-called universal nature may extend
to elements at all three \rpro\ peaks
(i.e., Se, Te, Os, Pt),
regardless of the variations that may occur 
between the first and second peaks.

We quantify the difference between the lighter and heavier
elements in our sample
by calculating the dispersions in the [Sr/Ba], [Sr/Eu],
[Ba/Eu], and [Eu/Yb] ratios.
The [Sr/Ba] and [Sr/Eu] ratios have 
dispersions of 0.29 and 0.27~dex, respectively.
The [Ba/Eu] and [Eu/Yb] ratios, however, have
dispersions of only 0.19 and 0.16~dex.
The latter two values are consistent with the
measurement uncertainties listed in Table~\ref{atmtab}.
This indicates that a real dispersion exists
between the lighter and heavier elements,
even among members of the class of highly-\rpro-enhanced stars.
The small dispersion in the [Ba/Yb] ratios (0.18~dex)
also suggests that the truncated \rpro\ \citep{boyd12}
did not affect the nucleosynthesis of the heavy elements
found in these stars.

In the solar system, approximately 94--98~per cent of the Eu is predicted
to have originated via \rpro\ nucleosynthesis, 
and the remainder is attributed to \spro\ nucleosynthesis
(e.g., \citealt{sneden08,bisterzo11}).
In contrast, only about 11--15~per cent of
Ba in the solar system
is predicted to have originated via \rpro\ nucleosynthesis.
Thus the [Ba/Eu] ratio is
commonly used to quantify the relative 
$r$- and \spro\ contributions to a given star or ensemble of stars.
The mean [Ba/Eu] ratio for all 13~stars in our sample is
$-$0.71~$\pm$~0.05 ($\sigma =$~0.19).
This compares well with other recent estimates by various methods
(e.g., 
$-$0.70, \citealt{arlandini99};
$-$0.81, \citealt{burris00};
$-$0.65, \citealt{sneden09};
$-$0.78, \citealt{mashonkina14}).

\section{Strongly $r$-process Enhanced Stars on the 
Subgiant and Horizontal Branches}
\label{hb}

Highly-\rpro-enhanced stars have
now been identified across a broad range of evolutionary states
that low-mass stars experience.
The first of these such stars identified, \rgb,
was a red giant, and for the following 15 years 
all other members of the class of highly-\rpro-enhanced
stars identified were also red giants.
\citet{aoki10} identified the first 
highly-\rpro-enhanced star on the main sequence,
\mbox{SDSS~J235718.91$-$005247.8}.
Our study has identified six new highly-\rpro-enhanced 
subgiants and three such RHB stars.

Highly-\rpro-enhanced subgiants have been found before
by the Hamburg/ESO R-process Enhanced Star Survey \citep{barklem05},
but the stars' relatively weak lines and 
the moderate spectral resolution used by \citeauthor{barklem05}\ 
limited their ability to derive abundances
of many \ncap\ elements.
\citet{preston06} also studied the same three RHB 
stars that we have analyzed, but the focus of that study was not
on deriving large numbers of \ncap\ elements to study
the abundances patterns in detail.

The unease among practitioners in the field 
that highly-\rpro-enhanced
stars are found almost exclusively among 
giants (e.g., \citealt{sneden08})
is put to rest.
The bias is observational.
Our results and those of \citet{aoki10} 
demonstrate that this phenomenon is
not limited to red giants.

\section{The Frequency of Carbon-Enhanced Metal-poor Stars with
High Levels of $r$-process Enhancement}
\label{carbon}

Enhanced [C/Fe] and [N/Fe] ratios were found in the 
first highly-\rpro-enhanced star discovered,
\rgb\ 
([C/Fe]~$= +$0.88, [N/Fe]~$= +$1.01;
\citealt{sneden96,sneden03}).
This raised the question of whether the 
C- and N-enhancement were related to the
\rpro\ enhancement.
\rgb\ is included in our sample. 

\sgc\ is also highly-enhanced in C and N
([C/Fe]~$= +$1.78, [N/Fe]~$= +$2.05).
\sgb\ exhibits modest C and N enhancement
([C/Fe]~$= +$0.69, [N/Fe]~$= +$0.49),
and C is also modestly enhanced in \sgf\
([C/Fe]~$= +$0.58).
All three of these stars are subgiants.
Detections of the CH, NH, or CN bands in other highly-\rpro-enhanced 
stars indicate that [C/Fe] and [N/Fe] are not enhanced.
Four subgiants and all three RHB stars
yield only uninteresting upper limits on [N/Fe], 
but the upper limits derived from the non-detection of CH
indicate that [C/Fe]~$< +$1.0 in each of the RHB stars.

The fraction of carbon-enhanced metal-poor stars is known to
increase with decreasing metallicity,
and these stars typically constitute $\approx$~7--32~per cent 
of local stellar samples with [Fe/H]~$< -$2.0
\citep{
beers92,
norris97,
beers05,
cohen05,cohen13,
frebel06,
lucatello06,
lai07,
placco11,
carollo12,carollo14,
lee13,
yong13}.
These estimates range by a factor of several because
of differing definitions of C enhancement,
metallicity ranges considered, sample sizes and selection techniques, and
median distance from the sun.
Some of these stars are found in binary systems
and presumably acquired their C enhancement 
through mass transfer from a more evolved companion.
Long-term velocity monitoring and
detailed studies of the \ncap\
abundance patterns indicate that this
explanation cannot be applied to other 
carbon-enhanced metal-poor stars
(e.g., \citealt{aoki02}, \citealt{ryan05}, \citealt{roederer14b}).

In our sample of highly-\rpro-enhanced stars,
finding 2 of 13~stars as carbon-enhanced (15~per cent)
is consistent with previous estimates.
There is no compelling reason to suspect that
the \rpro\ and C-enhancement are related.

\section{Light-Element Abundance Signatures Related to 
High Levels of $r$-process Enhancement}
\label{light}

The large, homogeneous dataset of \citet{roederer14}
affords a unique opportunity to 
compare abundances in individual stars to 
abundances in large samples of stars with similar stellar parameters.
These comparisons will be minimally affected by 
uncertainties in the stellar parameters, non-LTE effects,
poorly-known atomic data, etc.
Subtle but significant outlying abundance ratios
may be identified this way.
We use this technique to identify any
abundance signatures among the light elements,
those with 6~$\leq Z \leq$~30,
that are peculiar to the highly-\rpro-enhanced stars
relative to normal metal-poor stars.
Any peculiar abundance signatures found
could be associated with the supernovae or other
sites capable of generating large amounts of elements
beyond the second \rpro\ peak.

We perform this test for each of the 13 stars in our sample.
We begin by identifying a comparison sample of
stars for each \rpro-enhanced star
that are in the same evolutionary state,
have \teff\ within $\pm$~200~K, and
are within $\pm$~0.3~dex of the metallicity of the
\rpro-enhanced star.
We exclude other highly-\rpro-enhanced stars from the comparison samples.
We also exclude stars whose heavy elements indicate the presence
of a large amount of \spro\ material, since the
present-day composition of these stars may not 
accurately reflect the composition of their birth clouds.
The comparison samples typically include 20--30~stars
from \citet{roederer14}, but there as many as 61~stars
or as few as 6~stars for comparison in some cases.

\begin{figure*}
\begin{center}
\includegraphics[angle=0,width=2.9in]{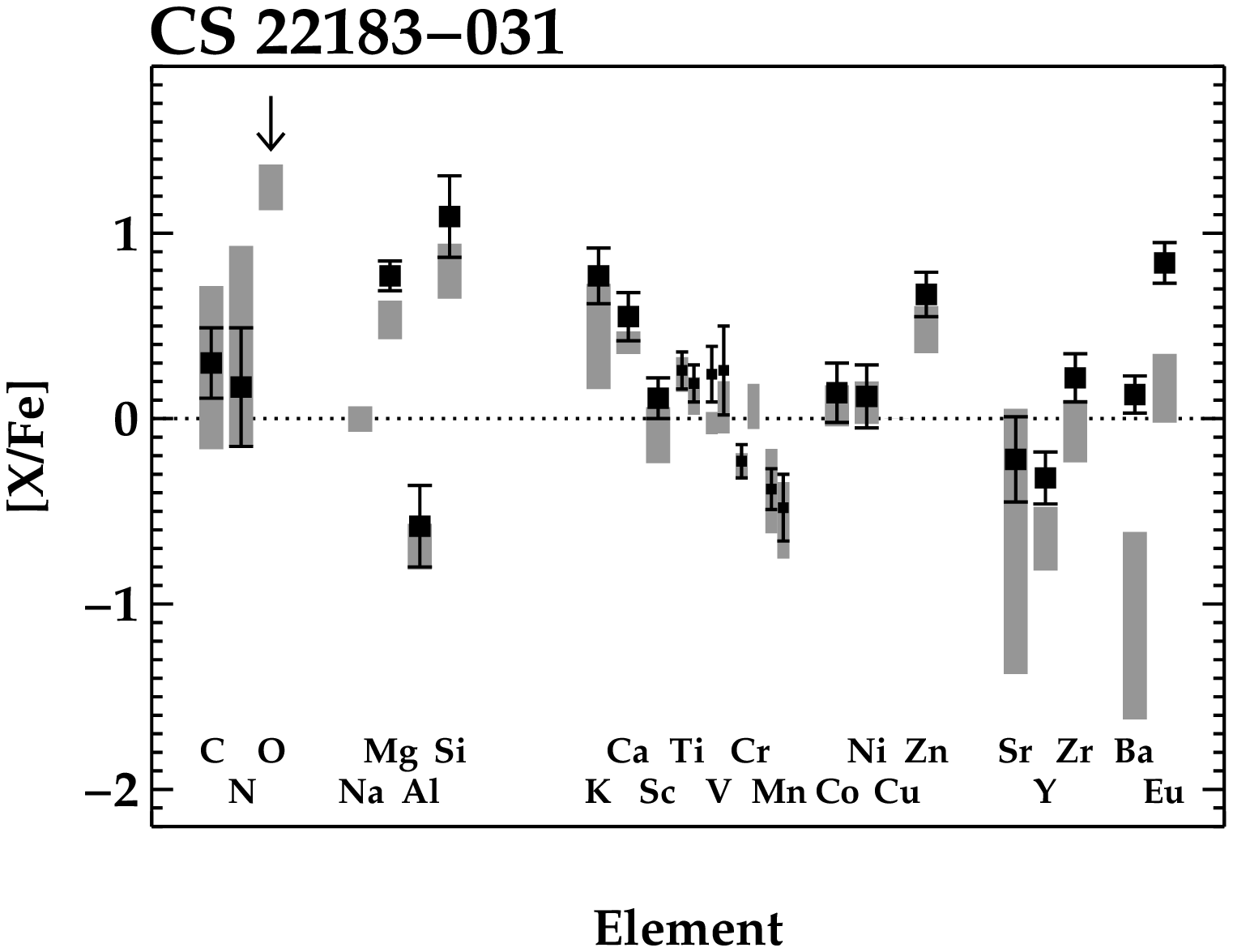}
\hspace*{0.3in}
\includegraphics[angle=0,width=2.9in]{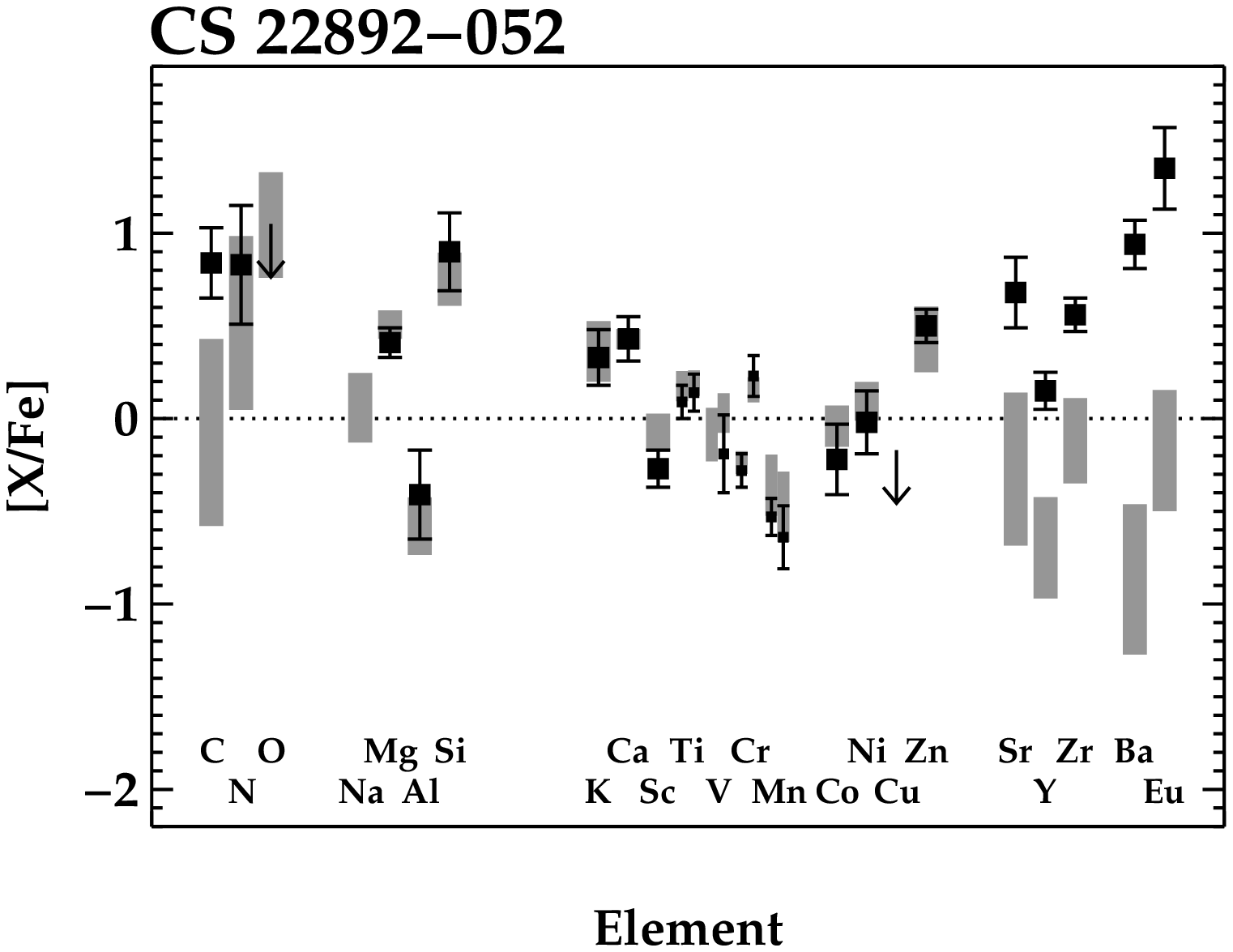} \\
\vspace*{0.3in}
\includegraphics[angle=0,width=2.9in]{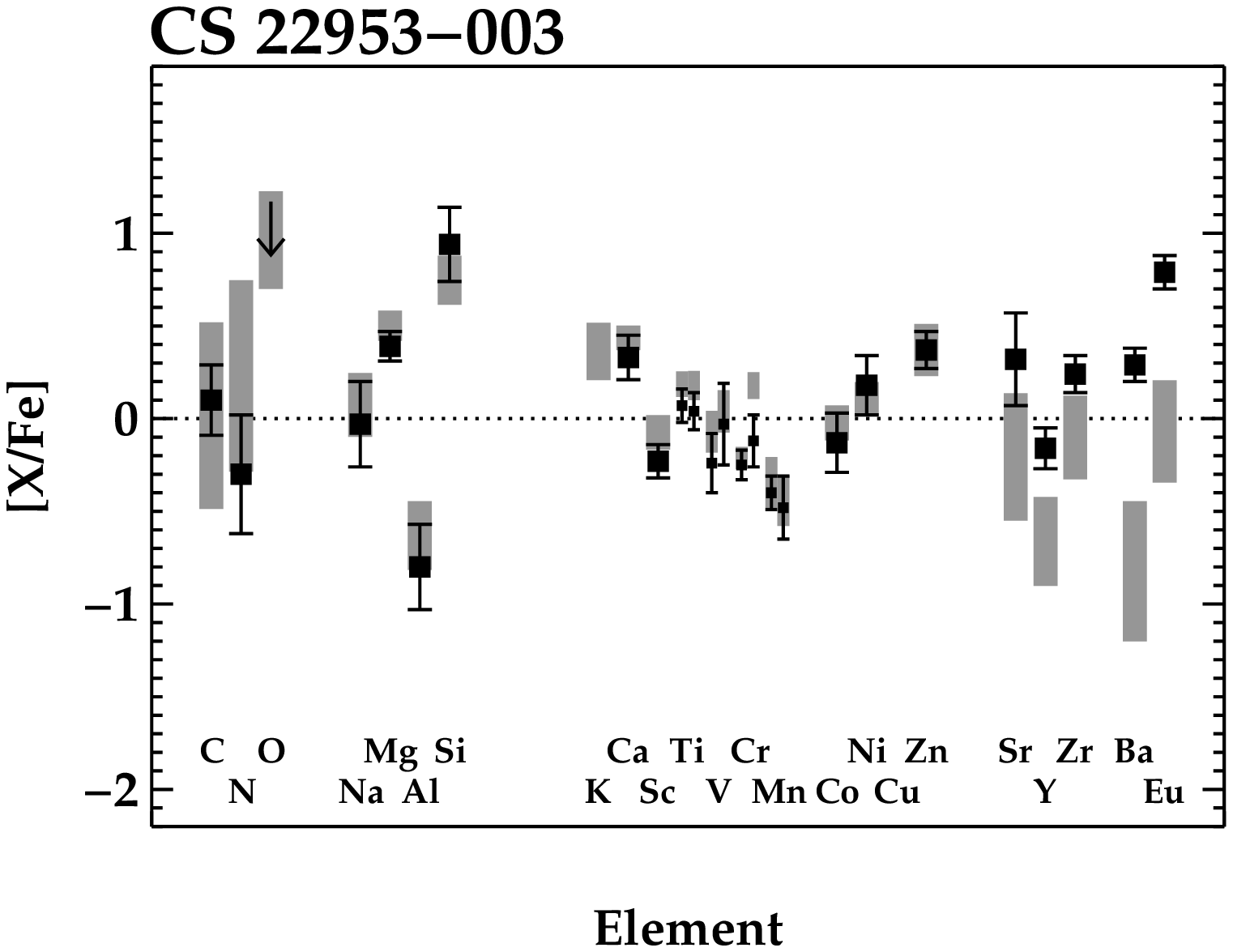} 
\hspace*{0.3in}
\includegraphics[angle=0,width=2.9in]{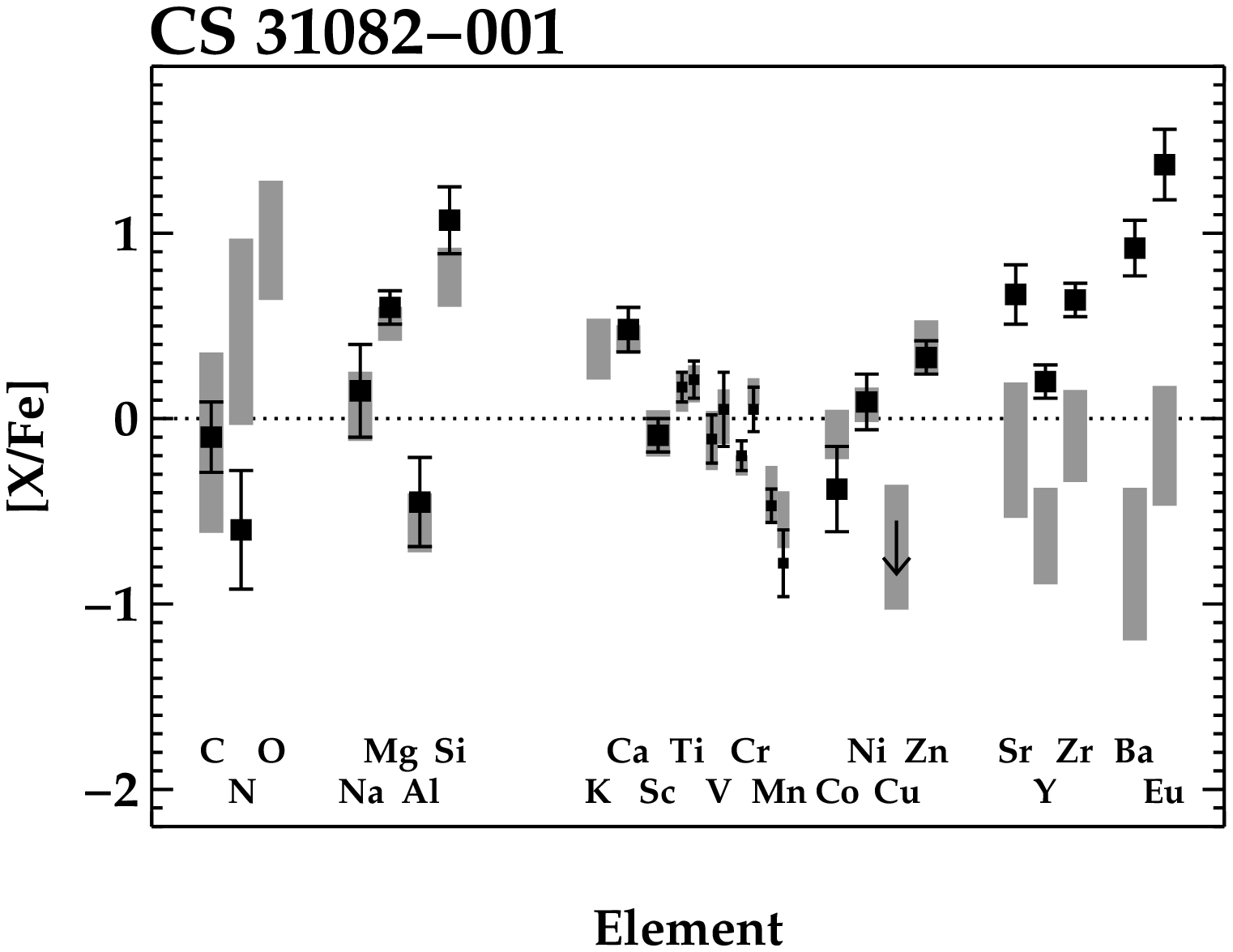} \\
\end{center}
\caption{
\label{comprgfig}
Comparison of abundances in the \rpro\ enhanced red giants
with the average abundances of other stars with
\teff\ within $\pm$~200~K and
 [Fe/H] within $\pm$~0.3~dex of each.
For
\mbox{CS~22183--031}, \mbox{CS~22892--052},
\mbox{CS~22953--003}, and \mbox{CS~31082--001},
the numbers of stars in the comparison samples are
12, 23, 30, and 23, 
respectively.
The comparison sample is shown by the shaded gray boxes,
representing the mean $\pm$~1$\sigma$ standard deviations.
The comparison sample is only
shown if it is derived from three or more stars.
Smaller symbols are shown for
Ti, V, Cr, and Mn
to accommodate ratios from both the neutral and ionized
states, which may differ.
The dotted line marks the solar ratios.
}
\end{figure*}

\begin{figure*}
\begin{center}
\includegraphics[angle=0,width=2.9in]{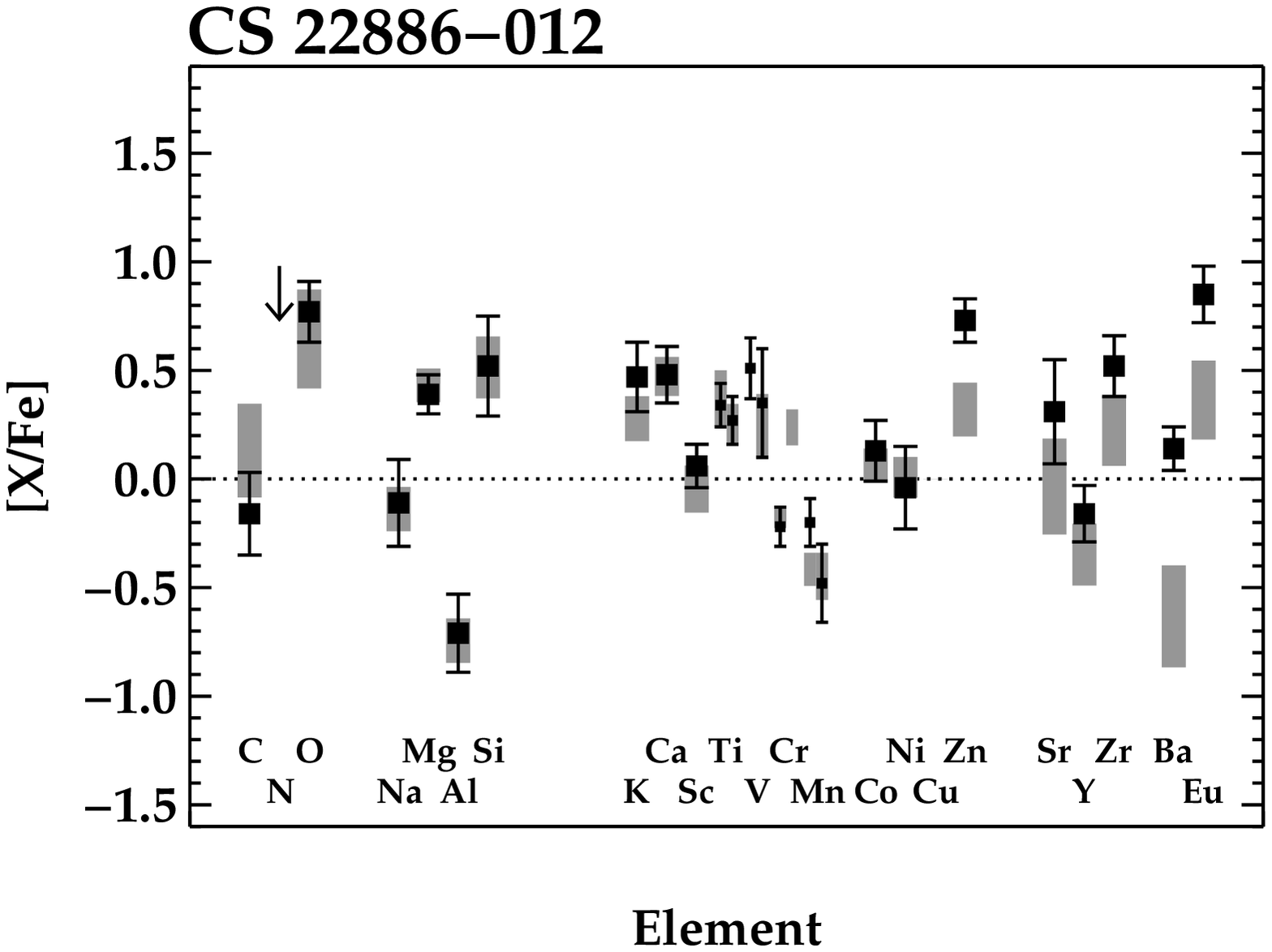}
\hspace*{0.3in}
\includegraphics[angle=0,width=2.9in]{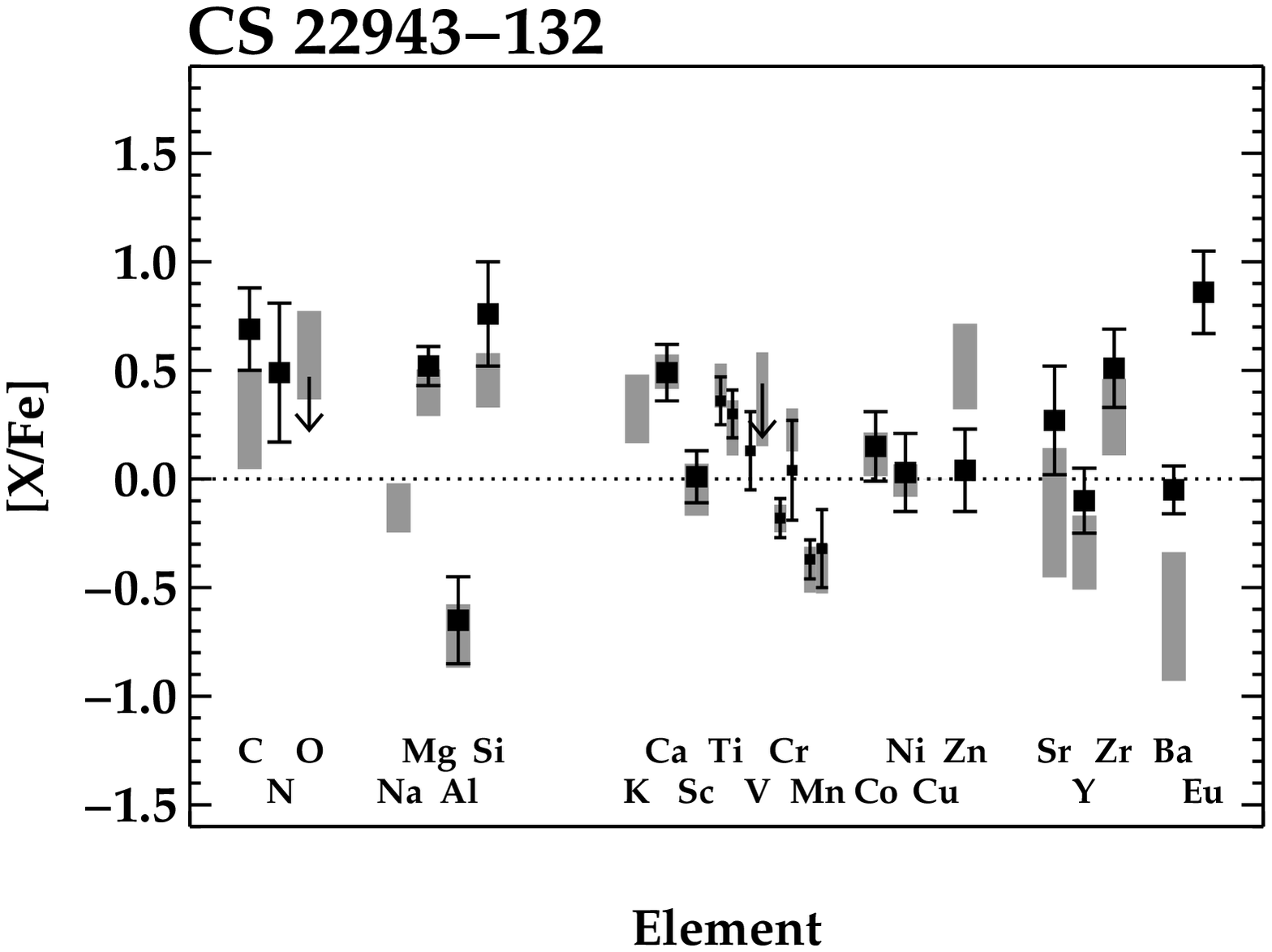} \\
\vspace*{0.3in}
\includegraphics[angle=0,width=2.9in]{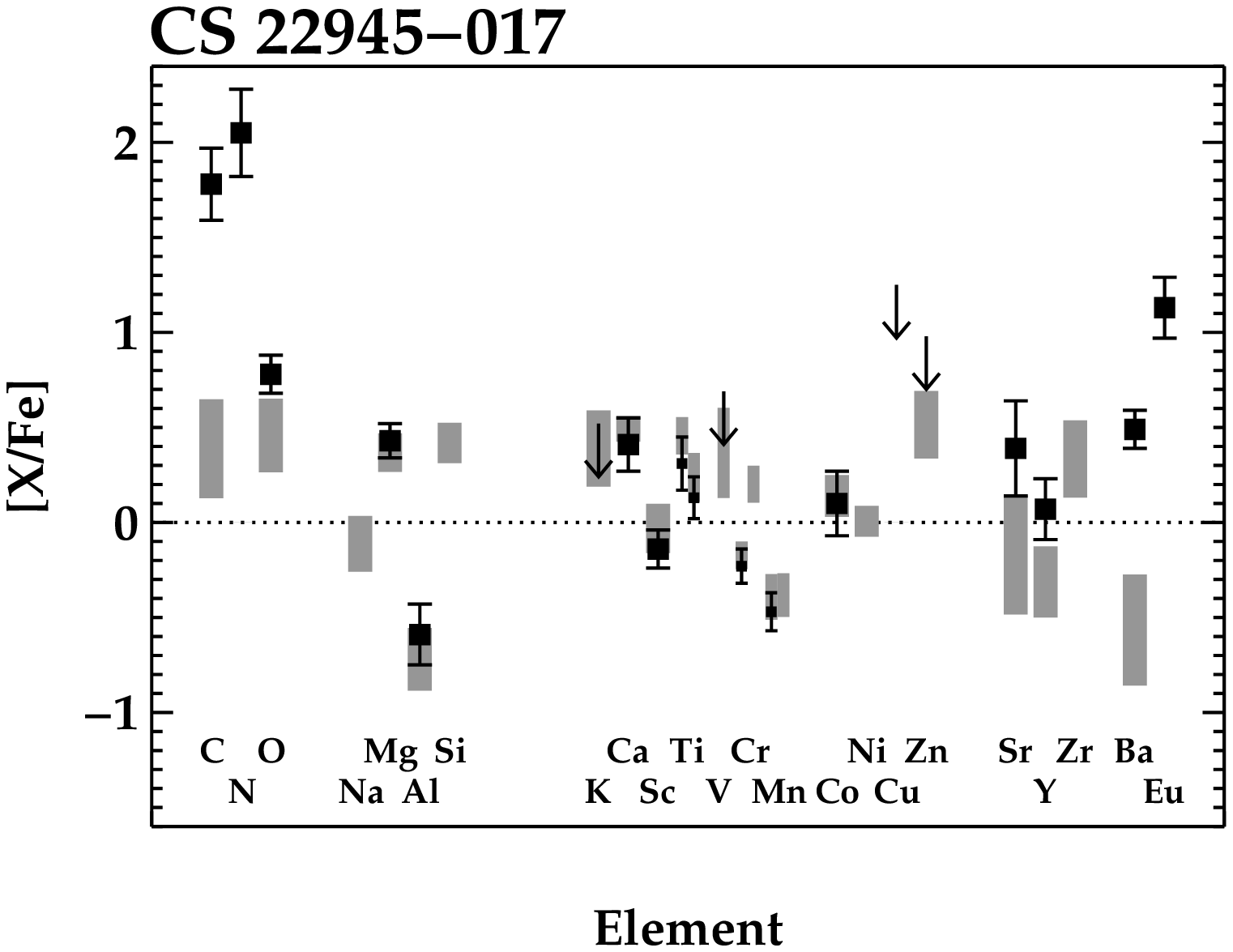}
\hspace*{0.3in}
\includegraphics[angle=0,width=2.9in]{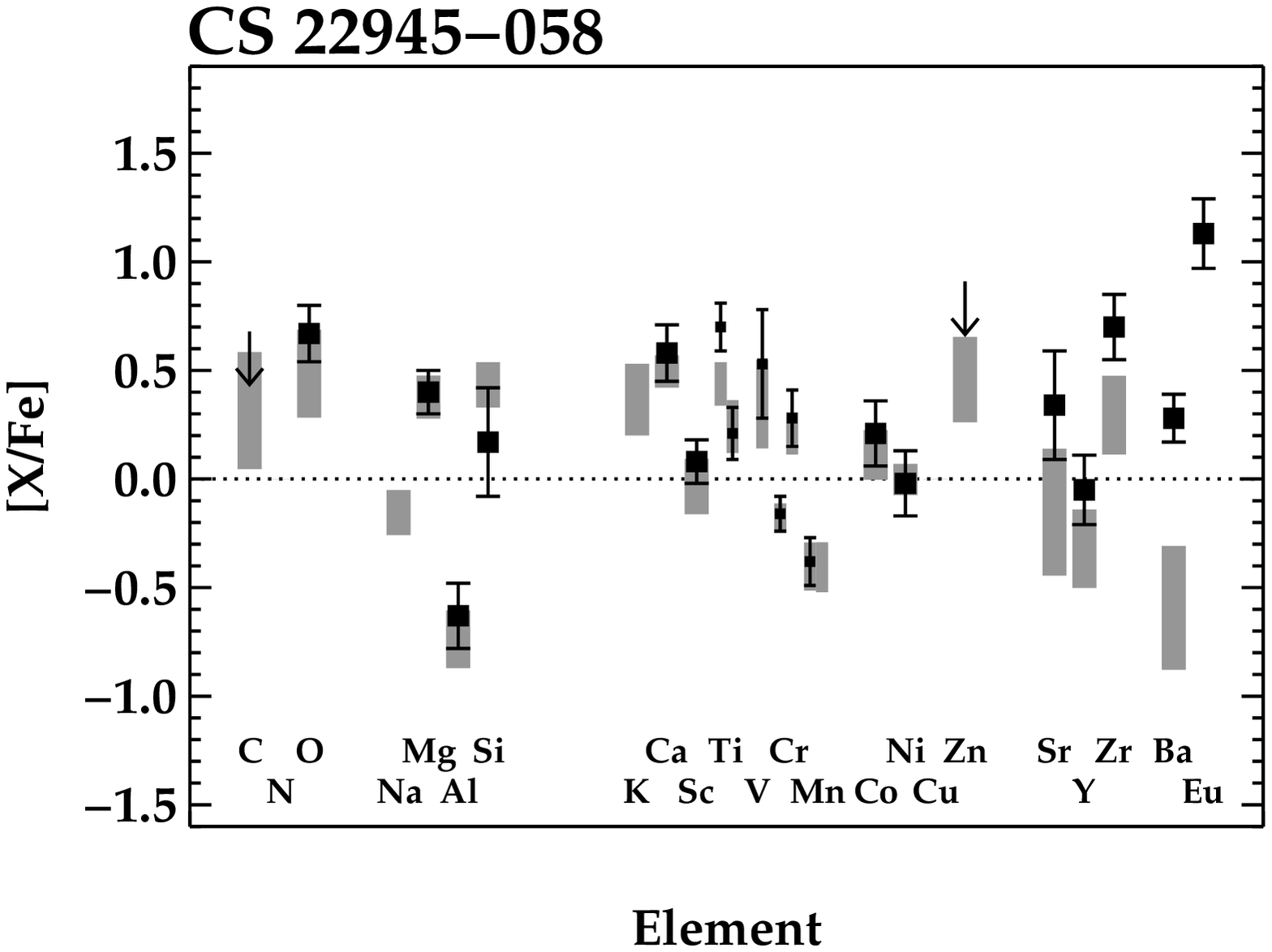} \\
\vspace*{0.3in}
\includegraphics[angle=0,width=2.9in]{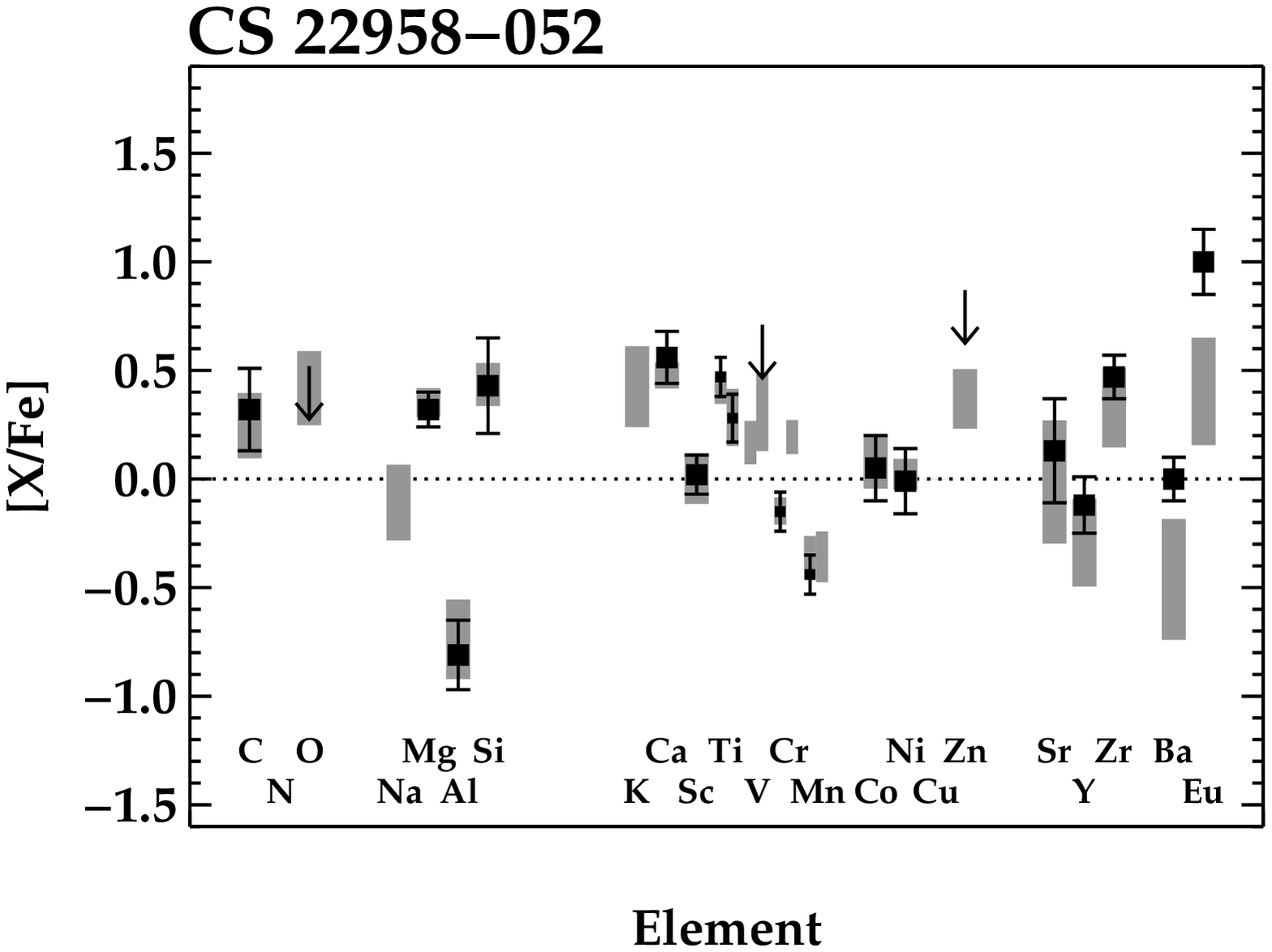}
\hspace*{0.3in}
\includegraphics[angle=0,width=2.9in]{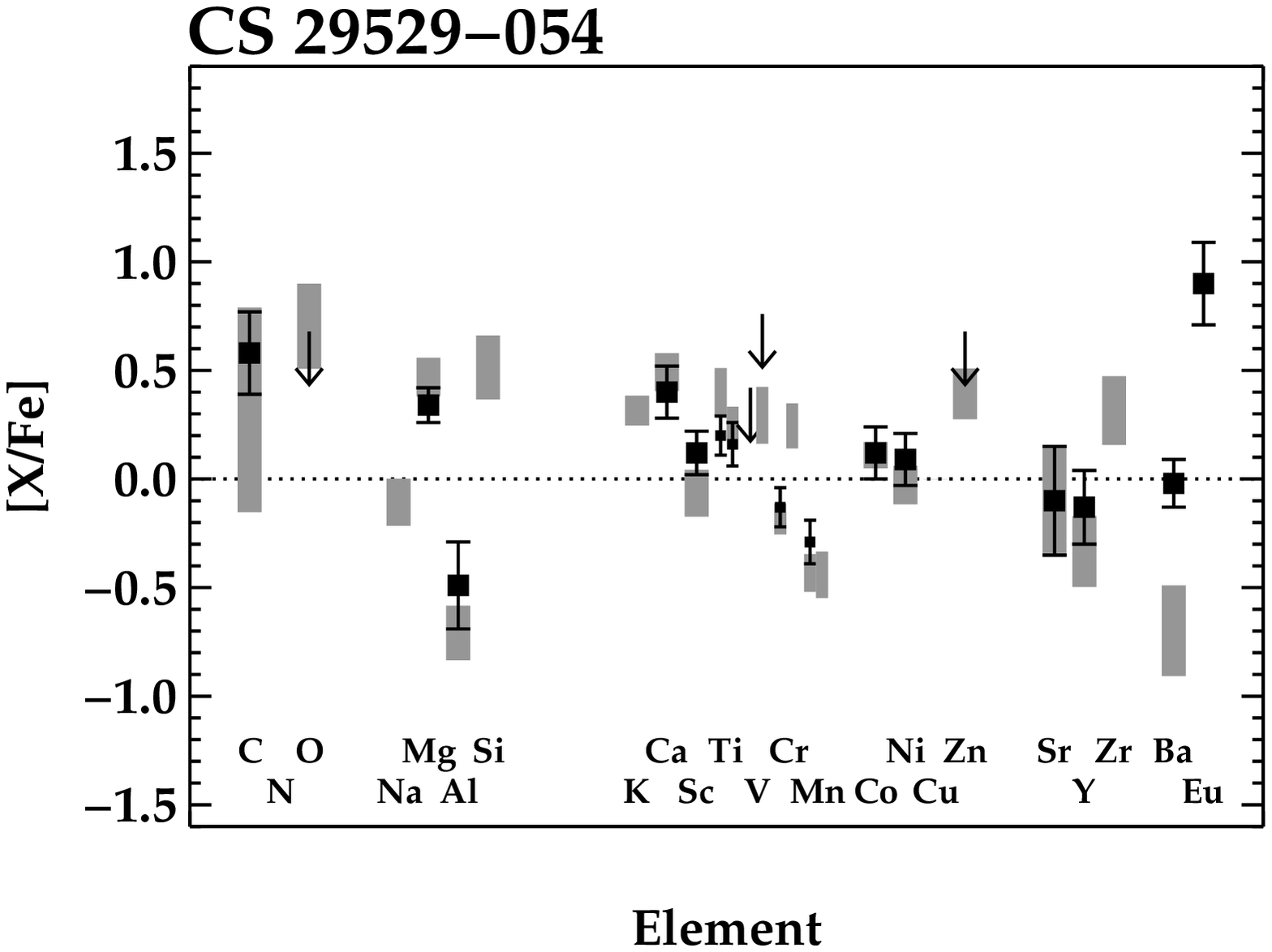} \\
\end{center}
\caption{
\label{compsgfig}
Comparison of abundances in the \rpro\ enhanced subgiants
with the average abundances of other stars with
\teff\ within $\pm$~200~K and
 [Fe/H] within $\pm$~0.3~dex of each.
For
\mbox{CS~22886--012}, \mbox{CS~22943--132}, \mbox{CS~22945--017}, 
\mbox{CS~22945--058}, \mbox{CS~22958--052}, and \mbox{CS~29529--054},
the numbers of stars in the comparison samples are
20, 50, 52, 61, 38, and 23,
respectively.
Symbols are the same as in Figure~\ref{comprgfig}.
}
\end{figure*}

\begin{figure}
\begin{center}
\includegraphics[angle=0,width=2.9in]{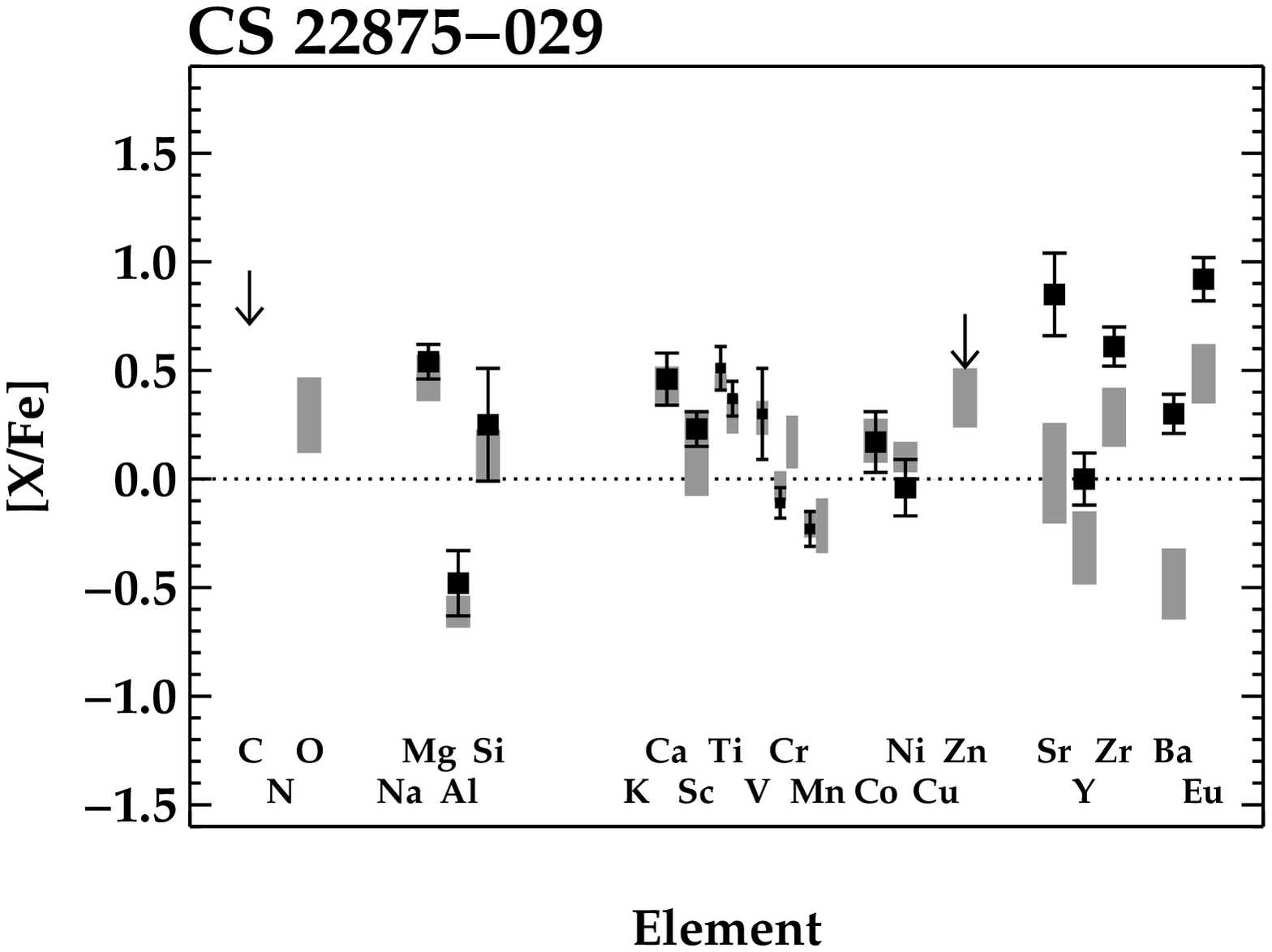} \\
\vspace*{0.3in}
\includegraphics[angle=0,width=2.9in]{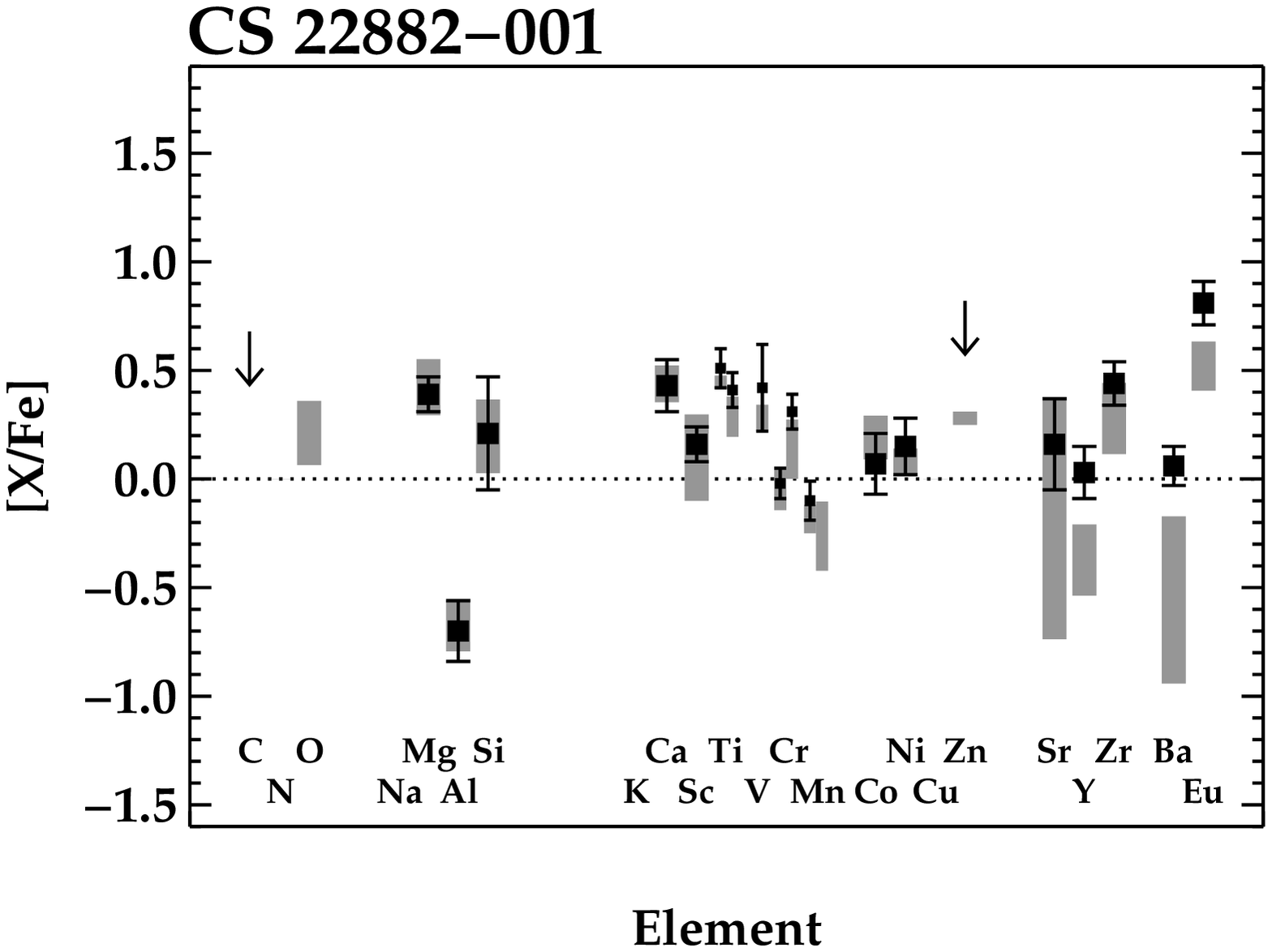} \\
\vspace*{0.3in}
\includegraphics[angle=0,width=2.9in]{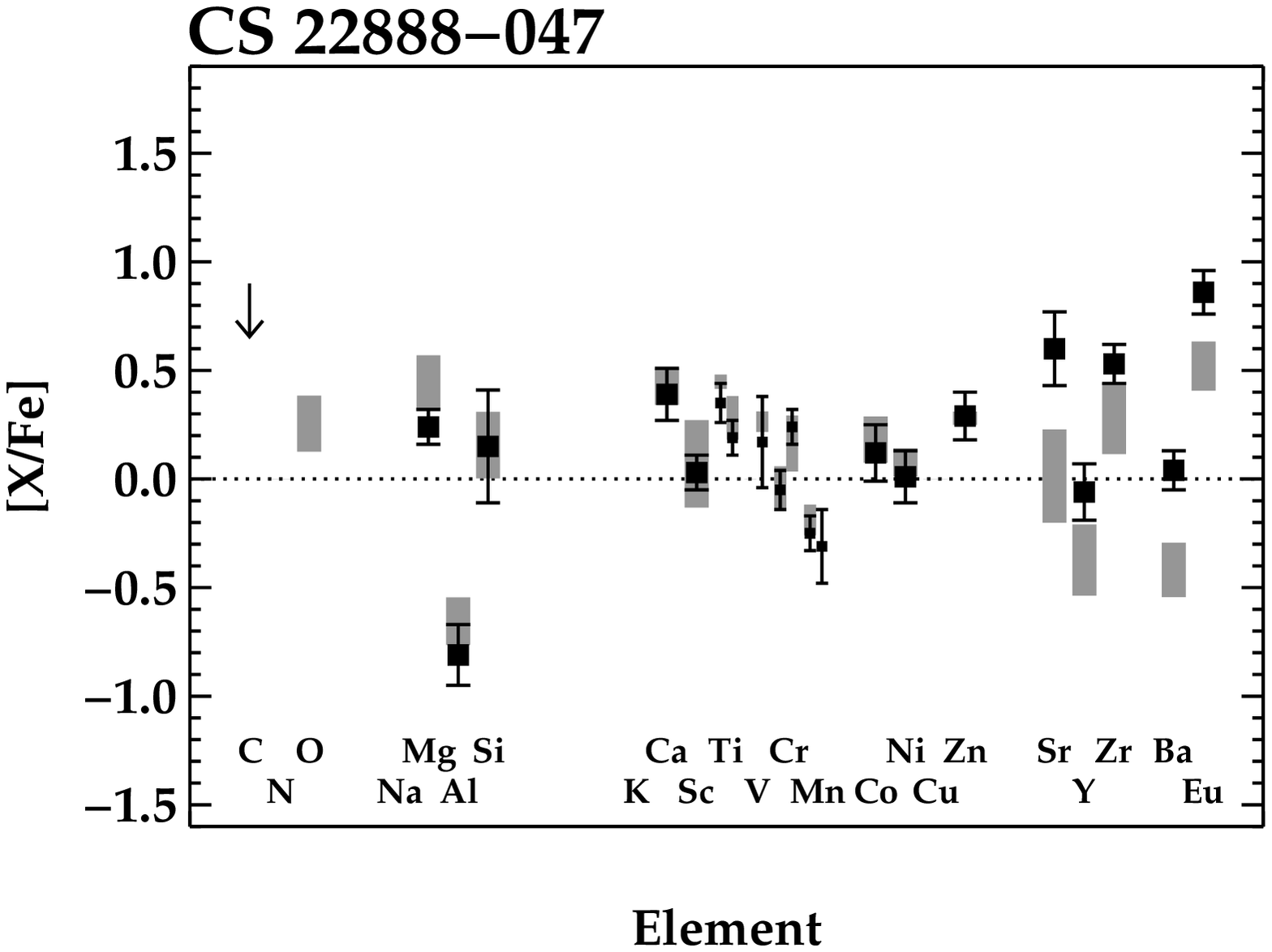} \\
\end{center}
\caption{
\label{comphbfig}
Comparison of abundances in the \rpro\ enhanced stars on the horizontal branch
with the average abundances of other stars with
\teff\ within $\pm$~200~K and
 [Fe/H] within $\pm$~0.3~dex of each.
For
\mbox{CS~22875--029}, \mbox{CS~22882--001}, and \mbox{CS~22888--047},
the numbers of stars in the comparison samples are
eight, seven, and six,
respectively.
Symbols are the same as in Figure~\ref{comprgfig}.
}
\end{figure}

Figures~\ref{comprgfig}--\ref{comphbfig}
illustrate the [X/Fe] ratios for each \rpro\ enhanced star.
The mean~$\pm$~1$\sigma$ standard deviation
of each [X/Fe] ratio for the comparison sample is marked
by the shaded box.  
The number of comparison stars is indicated in each figure caption.
These figures
illustrate that the light-element abundance ratios
of highly-\rpro-enhanced stars
rarely deviate by more than 1$\sigma$ from 
those of other metal-poor stars.
Like other metal-poor stars, the highly-\rpro-enhanced stars
show [$\alpha$/Fe] ratios
(where $\alpha$ indicates O, Mg, Si, Ca, and Ti)
enhanced by factors of a few relative to the solar ratios.
The iron-group elements are typically found in solar or sub-solar
ratios in both the \rpro\ enhanced stars and the normal metal-poor stars.
By construction, the \ncap\ elements Sr, Y, Zr, Ba, and Eu
are significantly enhanced relative to the
comparison stars.

\begin{figure}
\begin{center}
\includegraphics[angle=0,width=3.35in]{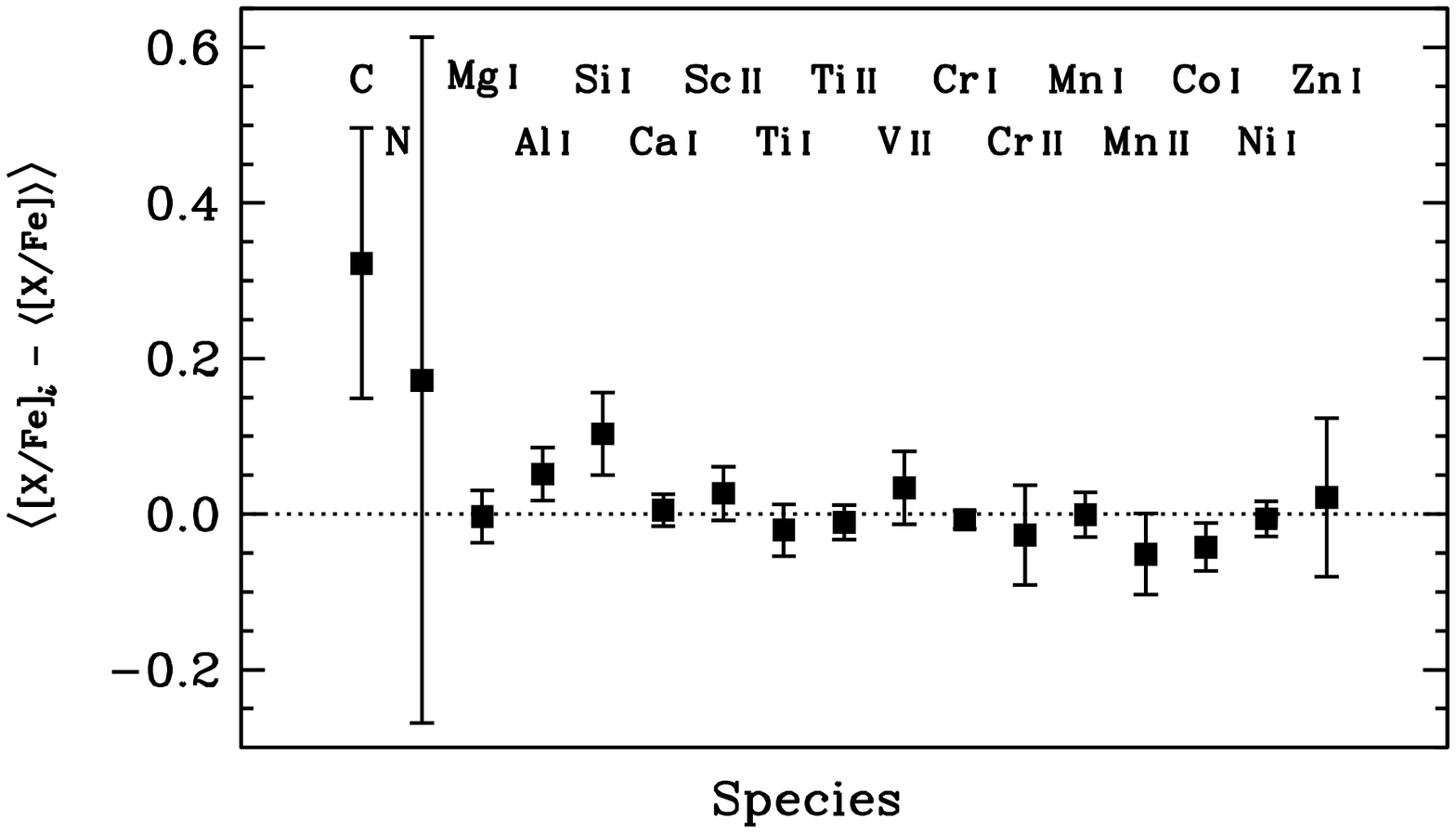}
\end{center}
\caption{
\label{meandelplot}
Mean residuals between each [X/Fe] ratio in each star
and its comparison sample.
The uncertainties correspond to the standard deviation of the mean.
Only comparisons between more than three stars are illustrated.
None of the mean residuals is significant
at the 2$\sigma$ level.
}
\end{figure}

Finally, we illustrate the mean deviations of each star
from its comparison sample in Figure~\ref{meandelplot}.
These values are listed in Table~\ref{meandeltab}.
The points in Figure~\ref{meandelplot} represent an unweighted
mean of [X/Fe]$_{i} - \langle$[X/Fe]$\rangle$ 
calculated for each element X in each \rpro\ enhanced star.
The error bars in Figure~\ref{meandelplot} represent the
standard deviation of the mean.
When multiple ionization states of an element are detected,
each ionization state is considered separately.
Species are only illustrated in Figure~\ref{meandelplot}
when the mean of the differences has been computed
from more than three stars.
None of the differences illustrated in Figure~\ref{meandelplot}
is significant at the 2$\sigma$ level or greater, and 
most light-element ratios in \rpro\ enhanced stars are 
consistent with the comparison samples at the 1$\sigma$ level.
The mean of the absolute values of these differences
for 15~[X/Fe] ratios from Mg to Zn is 
0.015~$\pm$~0.008~dex ($\sigma =$~0.030~dex),
or 3.5~per cent.

In summary, we find no compelling evidence that any of the
light-element ratios are significantly different in the
highly-\rpro-enhanced stars and metal-poor stars
without high levels of \rpro\ enhancement.
This complements results derived 
from a sample of seven
moderately-\rpro-enhanced stars
\citep{siqueiramello14}.
That study found no difference between the [X/Fe] ratios
(where X represents any of 13 elements between Mg and Ni)
among the \rpro-enhanced and unenhanced groups of stars.

\begin{table}
\begin{minipage}{2.9in}
\caption{Mean differences between $r$-process-enhanced stars and 
comparison samples
\label{meandeltab}}
\begin{tabular}{ccccc}
\hline
Ratio &
$\langle\Delta\rangle$ &
std.\ dev.\ &
std.\ err.\ & 
$N_{\rm stars}$ \\
\hline
{[C/Fe]}              & $+$0.322 & 0.52 & 0.17 &  9 \\
{[N/Fe]}              & $+$0.172 & 1.08 & 0.44 &  6 \\
{[Mg~\textsc{i}/Fe]}  & $-$0.003 & 0.12 & 0.03 & 13 \\
{[Al~\textsc{i}/Fe]}  & $+$0.052 & 0.12 & 0.03 & 13 \\
{[Si~\textsc{i}/Fe]}  & $+$0.103 & 0.18 & 0.05 & 11 \\
{[Ca~\textsc{i}/Fe]}  & $+$0.005 & 0.07 & 0.02 & 13 \\
{[Sc~\textsc{ii}/Fe]} & $+$0.027 & 0.12 & 0.03 & 13 \\
{[Ti~\textsc{i}/Fe]}  & $-$0.021 & 0.12 & 0.03 & 13 \\
{[Ti~\textsc{ii}/Fe]} & $-$0.010 & 0.08 & 0.02 & 13 \\
{[V~\textsc{ii}/Fe]}  & $+$0.034 & 0.14 & 0.05 &  9 \\
{[Cr~\textsc{i}/Fe]}  & $-$0.007 & 0.04 & 0.01 & 13 \\
{[Cr~\textsc{ii}/Fe]} & $-$0.027 & 0.17 & 0.06 &  7 \\
{[Mn~\textsc{i}/Fe]}  & $+$0.000 & 0.10 & 0.03 & 13 \\
{[Mn~\textsc{ii}/Fe]} & $-$0.051 & 0.13 & 0.05 &  6 \\
{[Co~\textsc{i}/Fe]}  & $-$0.042 & 0.11 & 0.03 & 13 \\
{[Ni~\textsc{i}/Fe]}  & $-$0.006 & 0.08 & 0.02 & 12 \\
{[Zn~\textsc{i}/Fe]}  & $+$0.021 & 0.27 & 0.10 &  7 \\
\hline
\end{tabular}
\end{minipage}
\end{table}

We conclude that the nucleosynthetic sites responsible 
for producing the large \rpro\ enhancement
did not produce any detectable light-element abundance signatures
unique from the core-collapse supernovae widely believed
to have produced the metals observed in the vast majority
of metal-poor stars.
This scenario may come about naturally by decoupling the
sites of large \rpro\ enhancement from 
core-collapse supernovae entirely;
e.g., if mergers of neutron stars are the 
source of this kind of \rpro\ enhancement.
Alternatively, if development of the conditions required
for this kind of \rpro\ nucleosynthesis 
is effectively decoupled from the
deep regions of supernovae where light-element production occurs,
this would also satisfy the observations.

A decoupling between light-element and \rpro\ nucleosynthesis
has been proposed previously 
(e.g., \citealt{wasserburg00}; \citealt{fields02}) 
to satisfy
the observed dispersion of [Eu/Fe] ratios at low metallicity
compared to, e.g., the [Mg/Fe] ratios,
as illustrated in Figure~14 of \citet{sneden08}.
Our results reaffirm this situation,
extend it to other elements, and
constrain the amount of variation
allowed in the individual light-element yields.

\section{Summary}
\label{summary}

We have examined the abundance patterns found in stars
with high levels of \rpro\ enrichment 
as indicated by their enhanced [Eu/Fe] ratios.
Our study differs from the \rpro\ survey of \citet{barklem05}
by deriving abundances from higher-resolution spectra
that cover a wider wavelength range.
Thirteen of the 313 metal-poor stars
analyzed by \citet{roederer14} 
are identified
as members of the class of $r$-II stars
after correcting for offsets in the derived [Eu/Fe] ratios
([Eu/Fe]~$\ga +$0.8).
Four of these stars are red giants 
whose high levels of \rpro\ enhancement 
were known previously.
We identify six new subgiants (including the MSTO phase)
and three new RHB
members of this class.
\citet{aoki10} also identified one such star on
the main sequence.
Once limited to the realm of red giants,
highly-\rpro-enhanced stars are now known all across
a broad range of stellar evolutionary states.

The first highly-\rpro-enhanced star discovered, \rgb,
was also enhanced in C and N,
which signaled a possible connection between
these chemical signatures.
Only 2 of the 13~stars in our sample (including \rgb)
are enhanced in
C and N, however,
offering no compelling evidence for such a connection.
We also find no compelling evidence to suggest that 
a disproportionately high fraction of
highly-\rpro-enhanced stars are members of binary or
multiple star systems, confirming the conclusions of 
a dedicated radial velocity survey by \citet{hansen11}.
The dispersion in the [Sr/Ba] and [Sr/Eu] ratios is
larger than the dispersion in the [Ba/Eu] or [Yb/Eu] ratios,
indicating that the robust pattern does not extend to the
elements between the first and second \rpro\ peaks
even within the class of highly-\rpro-enhanced stars.

We compare the light-element ($Z \leq$~30) abundance pattern
in each highly-\rpro-enhanced star with the light-element
abundance pattern in a comparison set of stars
that have similar stellar parameters 
but lack high levels of \rpro\ enhancement.
This test reveals no obvious light-element abundance signatures
that are unique to the highly-\rpro-enhanced stars.
The nucleosynthetic sites responsible 
for producing the large \rpro\ enhancement
apparently did not produce any detectable light-element abundance signatures
distinct from normal core-collapse supernovae 
responsible for much of the early metal production in the universe.

\begin{figure*}
\begin{center}
\includegraphics[angle=0,width=2.9in]{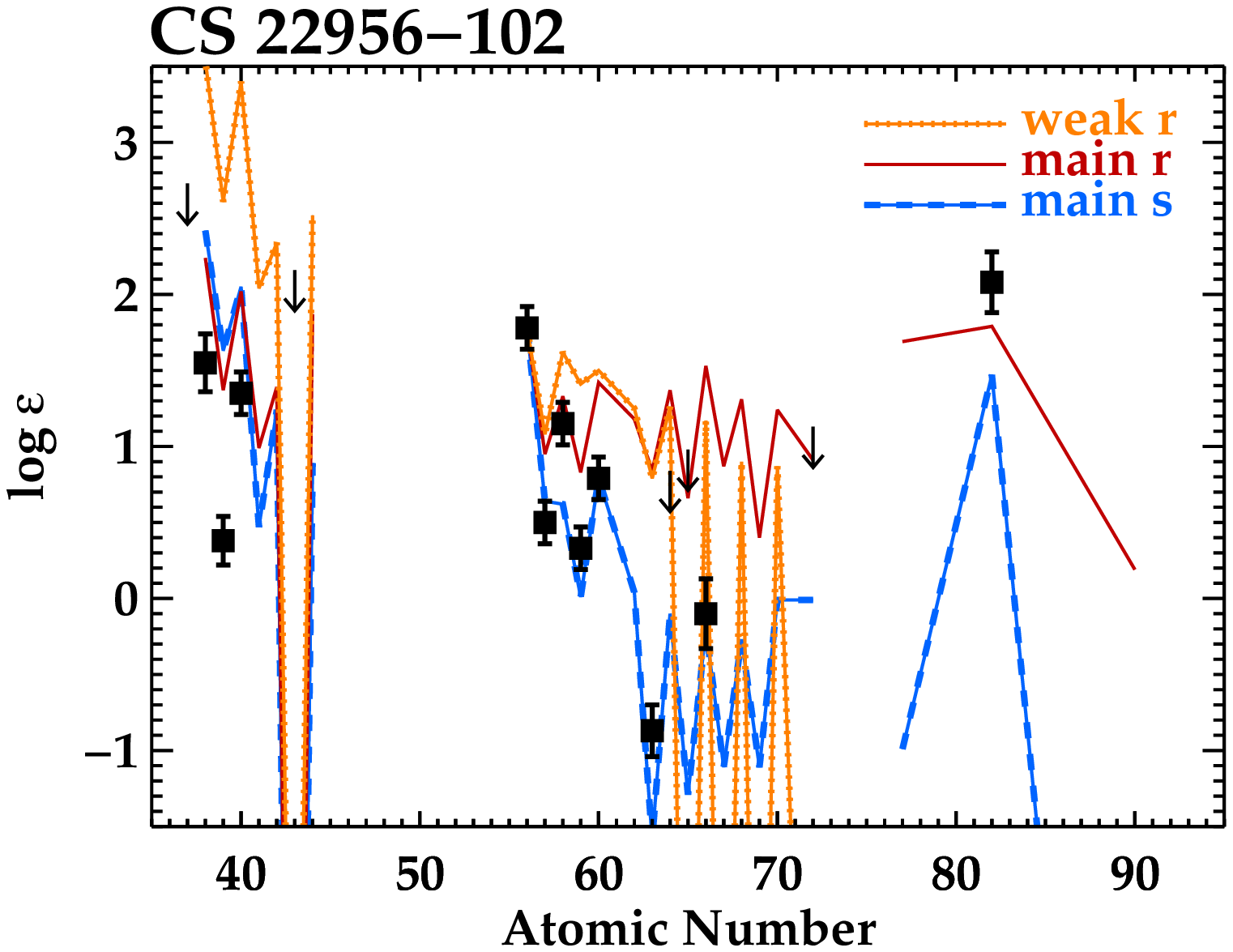}
\hspace*{0.3in}
\includegraphics[angle=0,width=2.9in]{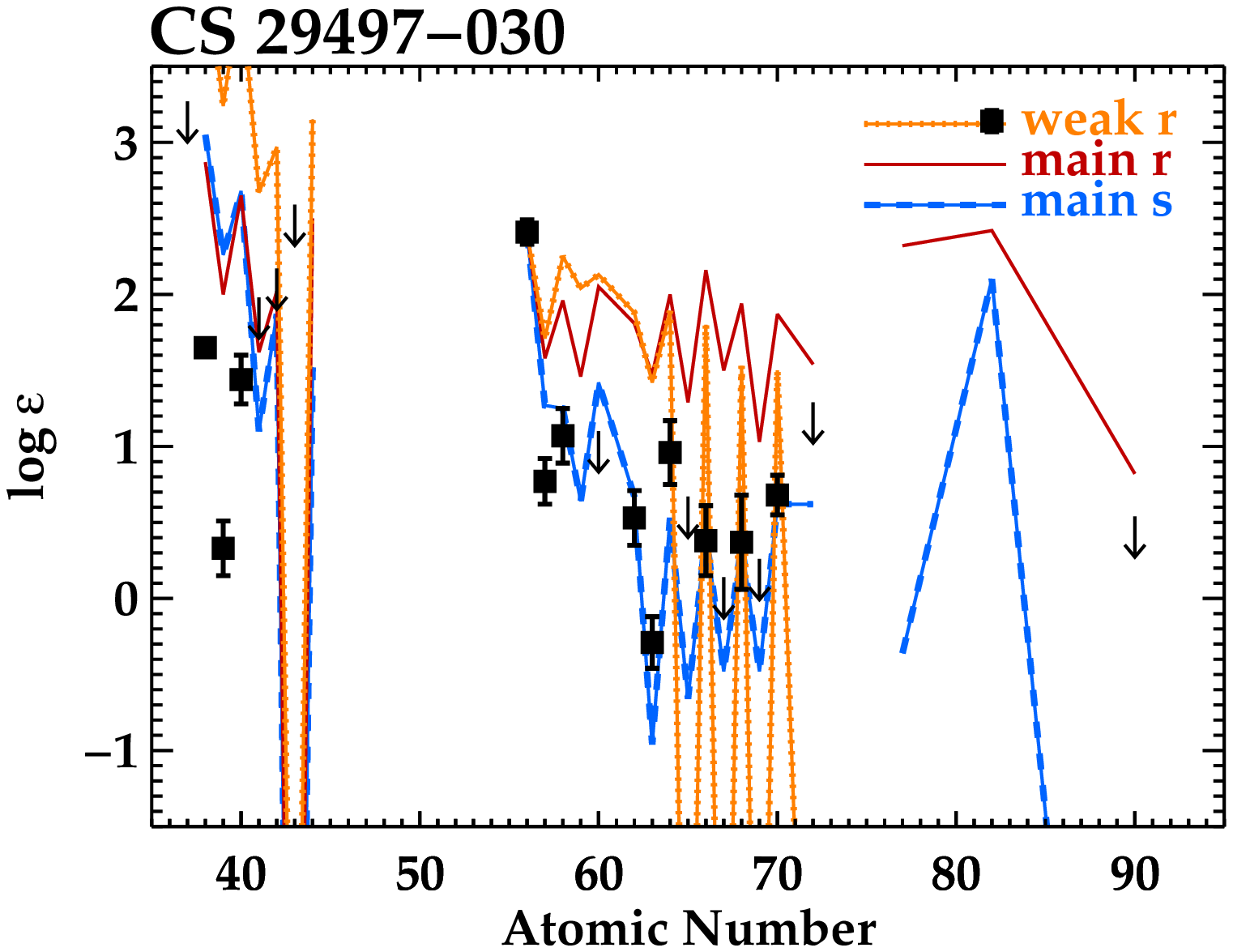} \\
\end{center}
\caption{
\label{sprofig}
The heavy element distributions in
\mbox{CS~22956--102} and \mbox{CS~29497--030}.
Symbols are the same as in Figure~\ref{rprorgfig}.
The curves have been normalized to the Ba abundance in each star.
}
\end{figure*}

\section*{Acknowledgments}

We thank the referee, T.\ Beers, for a prompt and helpful report.
This research has made use of NASA's 
Astrophysics Data System Bibliographic Services, 
the arXiv preprint server operated by Cornell University, 
the SIMBAD and VizieR databases hosted by the
Strasbourg Astronomical Data Center, and 
the Atomic Spectra Database hosted by
the National Institute of Standards and Technology. 
\textsc{iraf} is distributed by the National Optical Astronomy Observatories,
which are operated by the Association of Universities for Research
in Astronomy, Inc., under cooperative agreement with the National
Science Foundation.
C.~S.\ is supported by the U.S.\ National Science Foundation 
(grant AST~12-11585).


\appendix

\section{Two Stars with Europium Enhancements 
from $s$-process Nucleosynthesis}
\label{appendix}

\mbox{CS~29497--030}
has been identified previously using
traditional signatures of \spro\ enrichment from a binary companion
that passed through the thermally-pulsing asymptotic giant branch (TP-AGB)
phase of evolution:\ 
radial velocity variations,
a strong CH G band indicating substantial C enhancement,
and an \ncap\ abundance pattern indicative of \spro\ enhancement
\citep{preston00,sneden03b,ivans05}.
\citet{roederer14} derived
[C/Fe]~$= +$2.38,
[Ba/Fe]~$= +$2.75,
[Eu/Fe]~$= +$1.70, and
[Pb/Fe]~$= +$3.62
for 
\mbox{CS~29497--030}.
The \ncap\ abundance pattern of 
\mbox{CS~29497--030}
is illustrated in Figure~\ref{sprofig}.
Relative to the template \spro\ pattern shown for comparison,
elements at the first
\spro\ peak (38~$\leq Z \leq$~40) are underabundant
and lead ($Z =$~82) is overabundant.
This is explained by the high neutron-to-seed ratio
found in low-mass TP-AGB stars at low metallicity 
(e.g., \citealt{gallino98}).

\mbox{CS~22956--102}
also exhibits similar characteristics, as well as a 
weak but detectable C$_{2}$ band-head near 5165~\AA.
\citet{rossi05} identified 
\mbox{CS~22956--102}
as a carbon-enhanced metal-poor star, but \citet{roederer14}
published the first detailed abundance pattern derived
from high resolution spectroscopic observations.
\citeauthor{roederer14}\ also uncovered evidence of
radial velocity variations for this star.
\citeauthor{roederer14}\ derived
[C/Fe]~$= +$2.03,
[Ba/Fe]~$= +$1.81,
[Eu/Fe]~$= +$0.82, and
[Pb/Fe]~$= +$2.43 for
\mbox{CS~22956--102}.
The \ncap\ abundance pattern of 
\mbox{CS~22956--102}
is also illustrated in Figure~\ref{sprofig}.
Each of these stars exhibits unmistakable evidence
that the high levels of Eu enhancement
observed cannot be mainly attributed to \rpro\ nucleosynthesis.

\label{lastpage}

\end{document}